\documentclass[12pt]{article}
\usepackage[english]{babel}
\usepackage[latin1]{inputenc}
\usepackage{amsfonts,amssymb,amsmath, epsfig,hyperref}
\usepackage{color,graphicx,graphics,psfrag}
\usepackage{amsmath,amstext,amssymb,amsfonts, amscd}
\usepackage{caption}
\usepackage{multirow}
\usepackage{cases, subfigure, textcomp}

\def\Box{\leavevmode\vbox{\hrule
     \hbox{\vrule\kern4pt\vbox{\kern4pt}%
           \vrule}\hrule}}

\def\paragraph#1{{\bf #1\ }}

\newtheorem{lemma}{Lemma}[section]

\newtheorem{proposition}[lemma]{Proposition}

\newtheorem{remark}{Remark}[section]

\newcommand {\pd} {\partial}

\newcommand {\e} {\varepsilon}

\newcommand{\dd}{\,\mathrm{d}}

\title{An age-structured continuum model for myxobacteria} 
\author{Pierre Degond$^1$, Angelika Manhart$^2$, Hui Yu$^{3}$} 
\date{} 
\begin{document}

\maketitle

\begin{center}
1. Department of Mathematics, Imperial College London\\
London, SW7 2AZ, United Kingdom\\
pdegond@imperial.ac.uk 
\end{center}

\begin{center}
2. Courant Institute of Mathematical Sciences, \\
New York University, 251 Mercer Street, \\
New York, NY 10012, USA\\
angelika.manhart@cims.nyu.edu
\end{center}

\begin{center}
3. Yau Mathematical Sciences Center,\\
Jin Chun Yuan West Building, Tsinghua University, \\
100084 Beijing, China \\
huiyu@tsinghua.edu.cn
\end{center}

\begin{abstract}
Myxobacteria are social bacteria, that can glide in 2D and form counter-propa\-gating, interacting waves. Here we present a novel age-structured, continuous macroscopic model for the movement of myxobacteria. 
The derivation is based on microscopic interaction rules that can be formulated as a particle-based model and set within the SOH (Self-Organized Hydrodynamics) framework. 
The strength of this combined approach is that microscopic knowledge or data can be incorporated easily into the particle model, whilst the continuous model allows for easy numerical analysis of the different effects. However we found that the derived macroscopic model lacks a diffusion term in the density equations, which is necessary to control the number of waves, indicating that a higher order approximation during the derivation is crucial. Upon \textit{ad-hoc} addition of the diffusion term, we found very good agreement between the age-structured model and the biology. In particular we analyzed the influence of a refractory (insensitivity) period following a reversal of movement. 
Our analysis reveals that the refractory period is not necessary for wave formation, but essential to wave synchronization, indicating separate molecular mechanisms.
\end{abstract}


\vskip 0.4cm

%%%%%%%%%%%%%%%%%%%%%%%%%%%%%%%%%%%%%%%%%%%%%%%%%%%%%%%%%%%%%%%%%%%%%%%%%%%%%%%%%%%%%%%%%%%%%%%%
%%%%%%%%%%%%%%%%%%%%%%%%%%%%%%%%%%%%%%%%%%%%%%%%%%%%%%%%%%%%%%%%%%%%%%%%%%%%%%%%%%%%%%%%%%%%%%%%
\setcounter{equation}{0}
\section{Introduction}
\label{sec:intro}

Myxobacteria are a fascinating example for how simple cell-cell interaction rules can lead to emergent, collective behavior. These single-celled organisms have the ability to move on two dimensional surfaces and form large colonies. When \emph{swarming}, the colony exists as a rather uniform mono- or multi-layer of densely packed cells with single cells occasionally venturing away from the main swarm body. Myxobacterial swarms are predatory, searching and killing prey as a collective, which is one reason why these bacteria have often been called \emph{social} bacteria \cite{Dworkin1996}. Upon meeting prey, but also under starvation conditions the cells enter a \emph{ripple phase}, during which periodic density waves are formed \cite{Shimkets1990}. When two waves traveling in opposite directions collide, the waves appear to pass through each other unaffectedly. However, by tracking individual bacteria \cite{Sager1994,Welch2001} it was discovered that most cells in the wave crests in fact reverse their direction of movement, showing that the density waves are actually being reflected off each other. 
Myxobacteria reverse without turning, by internally exchanging the lagging and the leading pole. Isolated bacteria reverse spontaneously (on average every 10 min), however, their reversal rate increases as a response to higher densities of other bacteria around them. Although the precise function of rippling is not known, it often serves as a prelude and also overlaps with an \emph{aggregation phase}: in this developmental stage bacteria aggregate into several growing mounds which eventually rise out of the plane and form large three dimensional fruiting bodies. Both waves and aggregates are macroscopic structures with typical length scales of 100 $\mu m$, whereas individual bacteria are only a few microns long. Biologically this makes myxobacteria an interesting and suitable research object for understanding the development of multicellular cooperation, the basis of all complex life forms. Finally, the myxobacteria's unique metabolites have also rendered them an attractive source for potential new drugs \cite{reichenbach2001}.
\newline\par
The various social and cooperative behavior in observed myxobacterial colonies raises questions about the mechanisms of cell-cell communication. The most important mechanism responsible for inducing both ripple formation and aggregation has been found to be C-signaling: The C-factor is a 17-kD protein associated with the cell surface. It has been shown that direct cell-cell contact is necessary for C-signaling and that the exchange is facilitated via end-to-end contacts\cite{Kim1990}. Isolated cells exposed to purified C-factor show an increase in reversal frequency \cite{Sager1994}, suggesting that cell-to-cell contacts increase the probability for a cell to reverse. 
\newline\par
In this work we try to shed light on some of the questions associated with ripple formation: What primary effect of C-signaling causes a uniformly spread swarm to start forming ripples? Are density-dependent changes in reversal frequency enough to explain the formation of opposing, periodic wave trains?  One idea brought forward in \cite{Igoshin2001} and inspired by D. discoideum is that of an insensitivity or refractory period. Based on the observation that there seems to be a minimum time of around 40 sec \cite{Welch2001} between two reversals of the same bacterium, it is suggested that bacteria become insensitive to C-signaling immediately after they have reversed. Using mathematical modeling we show that a refractory period is not necessary for the formation of traveling waves as such, but rather for controlling the width of the waves as well as their wavelength. By analyzing the composition of waves in terms of insensitive and sensitive cells, we discover a possible mechanism how periodic waves are created and maintained in myxobacterial colonies.
\newline\par
While mutation experiments provide valuable insight, computational models offer a powerful alternative to test and analyze different mechanistic biological models. Detailed measurements and statistics about single cell behavior \cite{Sager1994,Welch2001} as well as mutation experiments provide the quantitative data necessary to formulate, parametrize and validate mathematical models. In many cases when modeling biological or physical systems one of the first modeling decisions is whether to use an individual- or particle-based model (IBM), in which the individual agents (in our case bacteria) interact by simple rules or to use a continuum model, in which the evolution of macroscopic quantities such as densities or mean directions are described by differential equations \cite{Manhart2016}. Advantages of IBMs are that they generally allow for an easy incorporation of biological knowledge or hypotheses and can deal with noise in a straightforward way. However the analysis of the model is often limited to running a large number of simulations and little mechanistic insight is gained. For (macroscopic) differential equations on the other hand  a large analytical toolbox ranging from asymptotic methods to linear stability analysis and bifurcation theory is available, which can produce precise results about the parameter dependence of solution behavior, etc. However for biological systems it is often difficult to derive continuous models, in many cases \textit{ad hoc} models are used in which some desired system behavior is already built into the derivation, thereby limiting the explanatory potential of the model.
\newline\par
The derivation of macroscopic models from IBMs of collective dynamics has been the subject of an intense literature with a particular focus on applications to  biology. This derivation proceeds through an intermediate modelling level called the kinetic or mean-field model \cite{blanchet2016topological,bolley2012mean,carlen2015boltzmann}. The derivation from kinetic to macroscopic models of collective dynamics faces the problem of the lack of conservation relations (such as the lack of momentum conservation, see e.g. the review in \cite{Vicsek_Zafeiris_PhysRep12}). A recent breakthrough is the so-called generalized collision invariant concept \cite{Degond2008,Frouvelle2012} subsequently developed in a variety of biological contexts (tissue self-organization \cite{Degond_Delebecque_Peurichard_2015}, flocking \cite{degond2016new}) as well as physical (micromagnetism \cite{Degond_Liu_2012}) or social (economics \cite{Degond_Liu_Ringhofer_2014}) contexts. Other models of collective dynamics can be found in \cite{Baskaran2008,bellomo2012,Bertin2009,Carrillo2009,Chate2008,Chuang2007,Eftimie2012,Lutscher2002,Mogilner1999,Peruani2011,Toner_Tu_1995}. In most works within the SOH framework, the derivation involves an expansion of the collision operator in terms of $\e$ -- the ratio between the microscopic and the macroscopic scale -- and a (formal) limit of $\e\rightarrow 0$. One effect of this approximation is the absence of any mass diffusion. For the myxobacteria model we found that this approach leads to a discrepancy between the particle model, in which the number of waves are controlled and the diffusion-free continuous model, in which they are not. Hence the $\mathcal{O}(\e)$ terms describing small, but finite effects seem to be crucial. In this work we use the IBM to estimate the correct size of the mass diffusion term, however future work will include a formal derivation using the Chapman-Enskog expansion (applied to an SOH model in \cite{degond2010}).
\newline\par
 In this paper, to model the refractory period between two reversals of myxobacteria, we use the concept of a local time that is reset to zero after each reversal. This idea is borrowed from similar ideas used in neuron dynamics (see e.g.~\cite{pakdaman2009dynamics}) itself being a variant of the renewal equation used to model the cell cycle \cite{gyllenberg1990nonlinear}.
\newline\par
The rest of the manuscript is organized as follows: In Sec.\,\ref{sec:model_presentation} we present a hierarchy of models starting with an Individual Based Model (IBM, Sec.\,\ref{ssec:particle}) and systematically deriving a PDE-based macroscopic model. The final, macroscopic 2-age model is presented in Sec.\,\ref{ssec:cont_2age}. In Sec.\,\ref{sec:numerics} we perform detailed simulations: first we validate the macroscopic model by comparison with the IBM (Sec.\,\ref{ssec:numerics_IBM}-\ref{ssec:add_diffusion}), then we further analyze the properties of the 2-age model, in particular with respect to experimental predictions and wave formation. Concluding remarks are found in Sec.\,\ref{sec:discussion}.

%%%%%%%%%%%%%%%%%%%%%%%%%
%%%%%%%%%%%%%%%%%%%%%%%%%
%%%%
%%%%  	MODEL PRESENTATION
%%%%
%%%%%%%%%%%%%%%%%%%%%%%%%
%%%%%%%%%%%%%%%%%%%%%%%%%

\begin{figure}[t!]
\centering
\includegraphics[width=\textwidth]{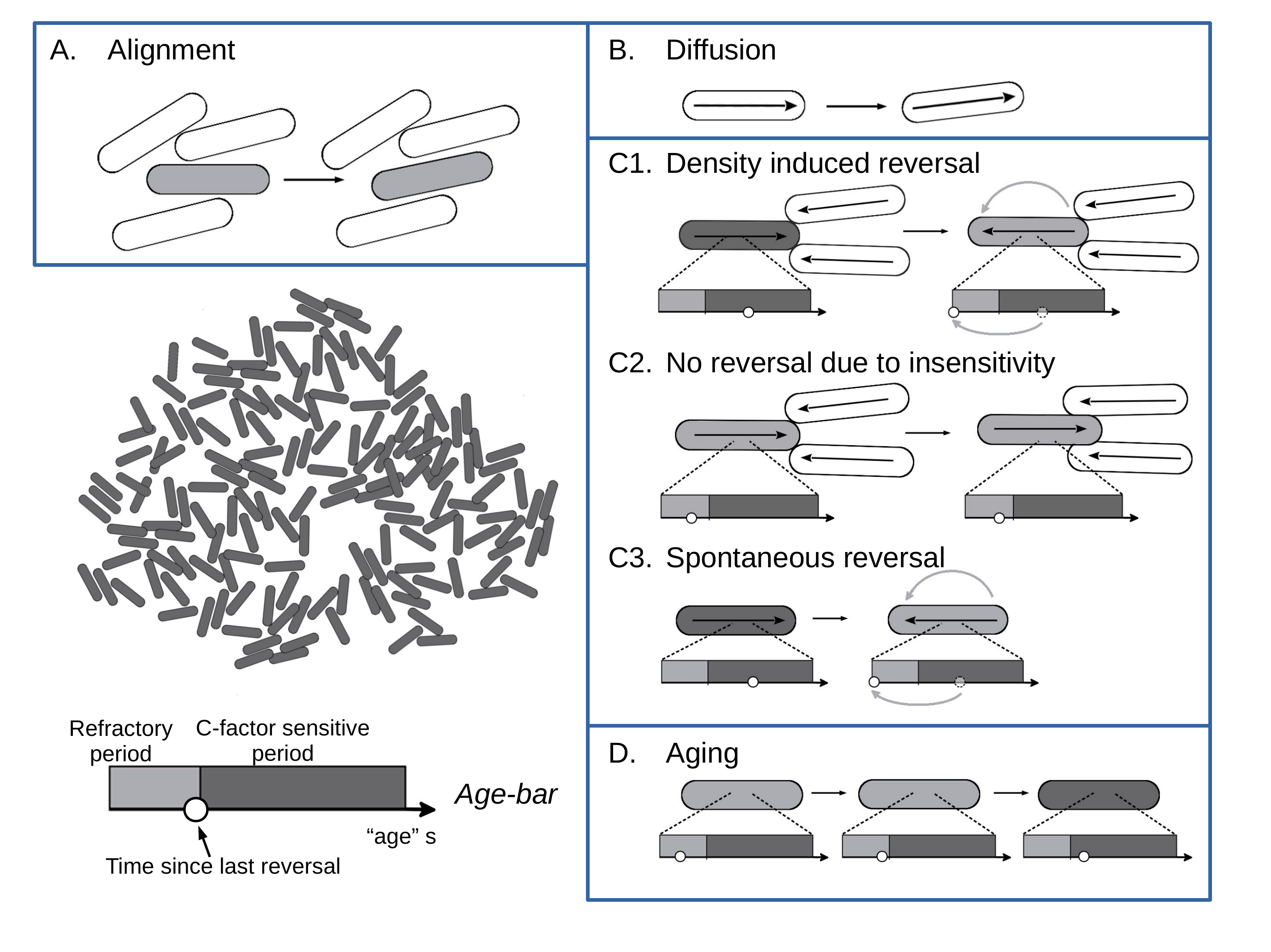}
\caption{Ingredients of the memory-dependent myxobacteria model. A: Nematic Alignment. The bacterium depicted in gray aligns nematically with its neighbors due to steric interactions and size exclusion effects. B: Angular Noise. The bacterium's orientation is subject to random fluctuation. C/D: Reversals and Aging. Bacteria can either be insensitive to C-signaling (light gray) or sensitive (dark grey). The age bar (explained in the lower, left corner) depicts the two periods and the bacterium's current state (white dot). Upon meeting oppositely moving bacteria, a bacterium can reverse, if sensitive to C-signaling (C1) or not, if in the refractory period (C2). They can also reverse spontaneously (C3). Insensitive bacteria age into a sensitive state (D) and reversals reset their age variable to zero (C1 and C3).}
\label{fig:IBM}
\end{figure}

\section{Model Presentation}
\label{sec:model_presentation}

We present a model hierarchy at the individual and macroscopic levels. 
We start with an individual-based model (Sec.\,\ref{ssec:particle}) describing the position and velocity of each bacterium as well as its internal biochemical age variable, which can be interpreted as memory of the bacterium. 
Here \emph{age} refers to the time passed since the last reversal. From there we systematically derive an age-structured continuous model for the macroscopic quantities, density and nematic mean direction (Sec.\,\ref{ssec:cont_agestructured}) where the age variable is still continuous. 
As a last step we discretize the age variable and assume only two groups or ages: being in the refractory period (i.e. being insensitive to C-signaling) or being sensitive to C-signaling. This results in a macroscopic 2-age model (Sec.\,\ref{ssec:cont_2age}), that forms the basis of the subsequent analysis.
\newline\par\noindent
\textbf{Model Assumptions.}  Fig.\,\ref{fig:IBM} shows the main model ingredients and assumptions:
\newline\par\noindent
\textbf{M1:} Bacteria move in 2D with constant speed, in the direction of their orientation. This orientation is subject to random noise (Fig.\,\ref{fig:IBM}.B).\newline
\textbf{M2:} Bacteria align nematically with other bacteria within their immediate vicinity (Fig.\,\ref{fig:IBM}.A).\newline
\textbf{M3:} Bacteria can reverse their orientation. Their reversal rate is a function of the local density of opposing bacteria (Fig.\,\ref{fig:IBM}.C1/C3).\newline
\textbf{M4:} After a reversal, bacteria go through a refratory period of fixed length, denoted by $T$, during which they cannot reverse (Fig.\,\ref{fig:IBM}.D/C2).
\newline\par
\noindent
\textbf{M2} can be interpreted as the effect of physical interactions between hard rods and models size exclusion effects and steric interactions. 
\textbf{M3} is a consequence of the contact-dependence of C-signaling, which we assume to act only over very short distances. 

%%%%
% PARTICLE MODEL
%%%%

\subsection{The Individual Based Model (IBM)}
\label{ssec:particle}

We describe the movement of $N$ individual myxobacteria. For $i\in \mathcal{N}:=\{1,..,N\}$, the $i$-th bacterium at time $t>0$ is characterized by its center of mass $X_i(t)\in \mathbb{R}^2$, its orientation angle $\Theta_i(t)\in[-\pi,\pi)$ (defined modulo $2\pi$) and an age variable $s_i(t)\geq 0$, related to the time since the bacterium's last reversal.
\noindent
\paragraph{Movement and Alignment.}
Bacteria move with constant speed $v_0>0$ in the direction $v(\Theta_i(t)) := (\cos\Theta_i(t), \sin\Theta_i(t))^T$. To model nematic alignment, we follow \cite{Degond2016} and assume that the following stochastic differential equations govern the evolution of $X_i(t)$ and $\Theta_i(t)$:
\begin{subequations}\label{IBM}
\begin{numcases}{}
\frac{\dd X_i}{\dd t} = v_0v(\Theta_i(t)), \\
\dd\Theta_i = -\nu{\rm Sign}\big(\cos(\Theta_i-\bar\Theta_i)\big)\sin(\Theta_i-\bar\Theta_i)\dd t 
-4D\cos(\Theta_i - \bar\Theta_i)\sin(\Theta_i - \bar\Theta_i)\dd t \notag \\
\hspace{1.25cm} + \sqrt{2D\cos^2(\Theta_i-\bar\Theta_i)}\dd B_t^i. \label{IBM_b}
\end{numcases}
\end{subequations}
\begin{remark}
\textit{Note that the term $4D\cos(\Theta_i - \bar\Theta_i)\sin(\Theta_i - \bar\Theta_i)\dd t$ presents a subtle difference to the model presented in \cite{Degond2016} and arises from a different interpretation of the stochastic differential equations (SDEs): Using the usual Ito convention, this term is necessary to be consistent with both the numerical implementation presented in App.\,\ref{app:numerics} and the Fokker-Planck equation derived in Sec.\,\ref{sec:derivation_SOH}.
In \cite{Degond2016} the SDEs were interpreted in the Backward Ito sense (also called isothermal convention \cite{Lancon2002}), and hence formulated without this term.}
\end{remark}
The parameter $\nu>0$ measures the alignment frequency to the local mean direction $\bar\Theta_i$. Bacteria either align with it (if $\cos(\Theta_i-\bar\Theta_i)>0$) or against it (if $\cos(\Theta_i-\bar\Theta_i)<0$). ${\rm Sign}$ is defined by ${\rm Sign(a)}=\pm 1$ for $\pm a>0$. The nematic mean direction $\bar\Theta_i$ can be understood as an average of lines going through each bacterium. It is defined by
\begin{align*}
\left(\begin{array}{c}
\cos(2\bar\Theta_i)\\
\sin(2\bar\Theta_i)\end{array}\right) 
= \frac{J_i}{|J_i|} \qquad \text{ with } \quad J_i = \sum_{k: |X_k-X_i|\leq R}
\left(\begin{array}{c}
\cos(2\Theta_k)\\\sin(2\Theta_k)\end{array}\right).
\end{align*}
$J_i$ represents the nematic mean current and the parameter $R>0$ specifies the interaction range of the alignment. Since it is an average of lines (as opposed to an average of angles) $\bar\Theta_i$ has to be understood modulo $\pi$ and we always choose $\bar\Theta_i\in[0,\pi)$. A more detailed discussion can be found in \cite{Degond2016}.
The angular noise is modeled by the stochastic process $\dd B_t^i$, describing independent Browinian motion of intensity $D\cos^2(\Theta_i-\bar\Theta_i)$, where $D>0$. The term $\cos^2(\Theta_i-\bar\Theta_i)$ aids the separation into two groups of bacteria traveling in opposite direction and is described in detail in \cite{Degond2016}. This concludes the modeling of assumptions \textbf{M1} and \textbf{M2}.\newline\par
\noindent
\paragraph{Reversals and Insensitivity.}
To model assumptions \textbf{M3} and \textbf{M4} we start by noting that bacteria can reverse their orientation, which changes $\Theta_i$ to $\Theta_i+\pi$. The reversal frequency depends on physical contact during which the signaling molecule C-factor is exchanged. We assume that immediately after a reversal, bacteria go through a refractory period of length $T$, during which they are insensitive to C-signaling. We therefore endow each bacterium with an \emph{age} variable $s_i(t)>0$ which measures the time elapsed since its last reversal, normalized by the refractory period $T$:
\begin{equation}
s_i(t) = \frac{t - \tau_i}{T},
\label{eq:age0}
\end{equation}
where $\tau_i$ records the time of the last reversal for the $i$-th particle and $s_i(t)$ is between $0$ and $\tau'_i - \tau_i$, where $\tau'_i$ is the instant of the next reversal. Therefore, the dynamics of $s_i$  between two instants of reversal is given by
\begin{align}
\left\{\begin{array}{ccl}
\frac{\dd s_i}{\dd t} &= \frac{1}{T}, &\text{ if } t > \tau_i \text{ and } t \text{ is not a reversal time,}\\
s_i &= 0, &\text{ if } t \text{ is a reversal time.}
\end{array}\right. \label{eq:age}
\end{align}
Eqs. (\ref{eq:age0}) and (\ref{eq:age}) are equivalent. While the former gives an explicit expression of the age variable $s_i$, the latter will be used in the formulation of the kinetic equation (\ref{eq:FP}) below. We model C-signaling sensitivity by a step function $\phi(s) \in \{0,1\}$:
\begin{equation}
\phi(s) = 
\left\{\begin{array}{ll} 0, &\text{ if } 0\leq s \leq 1 \text{ (refractory period)},\\ 
1, &\text{ if } s > 1 \text{ (C-factor sensitive period)}.\end{array}\right.
\label{eq:phi}
\end{equation}
To model \textbf{M3}, we assume that an individual bacterium's reversal rate is a function of the local density of bacteria oriented opposite to it. To that end, we define the local densities $\rho_\pm^i$ in the $i$-th bacterium's neighborhood $\mathcal{B}_i(X) = \{X: |X-X_i| \leq R\}$ as
\[
\rho^i_\pm = \frac{1}{|\mathcal{B}_i|}\text{Card}\{k\in \mathcal{N}\big|X_k\in \mathcal{B}_i\text{ and }
\pm \cos(\Theta_k-\bar\Theta_i)\geq 0 \},
\]
where `Card' is the cardinal of a set and $|\mathcal{B}_i|$ the area of $\mathcal{B}_i$. Since both are related to physical contact, we choose the interaction radii of density sensing and alignment to be equal, however in general they could be different.
The subscripts $\pm$ indicate whether the density refers to bacteria moving with ($+$) or against ($-$) the $i$-th particle. The reversal frequency as a function of density of opposing bacteria is denoted by $\lambda(\rho)$. Since higher concentration of opposing bacteria have been observed to cause more frequent reversals, we assume $\lambda(\rho)$ to be an increasing function of $\rho$. More discussion is provided in Sec.\,\ref{ssec:reversal}.
\newline\par
The total reversal function $\Lambda(\rho,s)$ takes into account both of the above factors, namely the density and the refractory period, and is defined by
\begin{equation}
\label{eqn:Lambda}
\Lambda(\rho,s) = \lambda(\rho)\phi(s).
\end{equation}
Finally, the probability that the $i$-th particle has reversed between $t$ and $t+\Delta t$ is modeled as a Poisson process:
\begin{align*}
\text{Pr}\{\Theta_i(t+\Delta t) = &\Theta_i(t) + \pi \}=\\
&\Lambda(\rho^i_{-{\rm Sign}\cos(\Theta_i-\bar\Theta_i)},s_i(t))\Delta t\, \text{exp}\big(-\Lambda(\rho^i_{-{\rm Sign}\cos(\Theta_i-\bar\Theta_i)},s_i(t))\Delta t\big).
\end{align*}

This completes the description of the Individual Based Model (IBM). Numerical results are presented in Sec.\,\ref{ssec:numerics_IBM}.

%%%%
% CONT. AGE
%%%%

\subsection{The Macroscopic Continuous-Age Model}
\label{ssec:cont_agestructured}

The IBM presented in Sec.\,\ref{ssec:particle} consists of $4N$ coupled stochastic differential equations for typical bacterial colony sizes of $N\approx 10^6$. 
We therefore derive a macroscopic model that consists of only three partial differential equations. 
The derivation strategy uses the \emph{Self-Organized Hydrodynamics} (SOH) framework, which allows a systematic derivation of hydrodynamic equations for particle systems that do not have enough conserved quantities, a common obstacle in biological systems. 
The derivation follows \cite{Degond2016} and is described in more detail below in Sec.\,\ref{sec:derivation_SOH}; however its structure can be summarized as follows. 
First a mean field model is derived, which leads to a Fokker-Planck equation for the 1-particle distribution function. A hydrodynamic scaling introduces a small parameter representing the difference in microscopic and macroscopic time and spatial scales. Taking this parameter to zero, one finds that the equilibrium distribution function is locally characterized by three quantities: The nematic mean direction $\bar\theta$, the density of particles aligned with it, $\rho_+$ and that of particles anti-aligned with it, $\rho_-$. The macroscopic model describes how these three quantities change in time and space and how the densities depend on the bacteria's biochemical age.

%%%%%%%%%
%  DERIVATION I: CONT. AGE
%%%%%%%%%

\subsubsection{Derivation of the Macroscopic Continuous-Age Model \eqref{eqn:macro_main}}
\label{sec:derivation_SOH}
The particle model presented in Sec.\,\ref{ssec:particle} serves as the starting point for the derivation. We largely follow \cite{Degond2016} and emphasize differences where appropriate.\newline\par
\noindent
\paragraph{Kinetic Equation.} Following the classical strategy for mean field models, presented e.g. in \cite{Degond2008}, we let the number of particles $N$ tend to infinity. Then the distribution function $f(x,\theta,s,t)$ satisfies the following Kolmogorov-Fokker-Planck type equation:
\begin{equation}
\partial_t f + v_0\nabla_x\cdot(v(\theta)f) + \frac{1}{T}\partial_s f = Q^R_\text{al}(f) + Q^R_\text{rev}(f).
\label{eq:FP}
\end{equation}
 We note in particular that the term $\frac{1}{T}\partial_s f$ comes from the age equation (\ref{eq:age}). 
We recall that $R$ denotes the interaction radius for both reversals and alignment. Here and in the following the superindex $R$ is used to emphasize the non-locality of the corresponding terms. The collision operator $Q^R_\text{al}$, caused by the alignment is almost identical to that derived in \cite{Degond2016} and is given by
\[
Q^R_\text{al}(f) = \partial_\theta\left[\nu{\rm Sign}\left(\cos(\theta-\bar\theta^R_f)\right)\sin(\theta-\bar\theta^R_f)f+D\cos^2(\theta-\bar\theta^R_f)\pd_\theta f\right],
\]
where the nematic mean direction $\bar\theta^R_f$ is defined by
\[
v(2\bar\theta^R_f(x, t)) = \frac{J^R_f(x,t)}{|J^R_f(x,t)|}, \quad 
J^R_f(x,t) = \int_0^\infty\int_{|x-y|\leq R}\int_{-\pi}^{\pi}v(2\theta)f(y, \theta, s, t)\dd y \dd\theta \dd s.
\]
Note that the definition of the mean nematic current $J_f^R$ requires integrating over all ages $s$, a small difference to the operator of the age-free model defined in \cite{Degond2008}, which reflects the fact that the biochemical age of a bacterium does not influence the alignment. The first term in $Q^R_\text{al}$ is a drift term in $\theta$ that moves the mass towards $\bar\theta^R_f$ and $\bar\theta^R_f-\pi$. The second term causes diffusion in $\theta$ with magnitude $D\cos^2(\theta-\bar\theta_f^R)$.
\newline
The new operator $Q^R_\text{rev}(f)$, describes the reversals and is defined as follows.
First we introduce two quantities $\sigma_f^{R,\pm}(x,t)$ representing the position-dependent, total densities of each group,
\begin{align*}
\sigma_f^{R,\pm}(x,t) = \frac{1}{\pi R^2}\int_0^\infty\int_{\pm\cos(\theta - \bar\theta^R_f)>0}\int_{|x-y|\leq R} f(y, \theta, s, t)\dd y \dd\theta \dd s.
\end{align*}
Note that these quantities are densities, i.e. number of particles per unit area, hence the factor $\pi R^2$ at the denominator, representing the area of the set $\{ y \, \, | \, \, |x-y|\leq R \}$. Then we have
\begin{align}
\label{eqn:QRrev}
Q^R_\text{rev}(f) = - \left[\Lambda(\sigma_f^{R,-},s)\chi_{\{\cos(\theta-\bar\theta^R_f)>0\}} + \Lambda(\sigma_f^{R,+},s)\chi_{\{\cos(\theta-\bar\theta^R_f)<0\}}\right]f(x,\theta,s,t),
\end{align}
where $\chi_{\mathcal{S}}$ is the characteristic function on the set $\mathcal{S}$ and $\Lambda(\sigma,s)$ is defined in \eqref{eqn:Lambda}. Eq. \eqref{eqn:QRrev} is nonzero only for $s>1$, since only then particles are sensitive to C-signaling and can reverse with a frequency that depends on the density of the opposing group. Note that \eqref{eqn:QRrev}  represents particles reversing \emph{away} from their group. Indeed, bacteria having reversed only enter the balance at age $s=0$. Away from $s=0$ there is no influx of (turned) bacteria, which explains why in \eqref{eqn:QRrev} the loss term appears alone. The contribution of bacteria \emph{having reversed} is accounted for as a boundary conditions at age $s=0$ given by
\begin{align}
f(x,\theta, 0, t) = T\!\!\int_0^\infty\!\!\Big(&\Lambda(\sigma_f^{R,-},s')\chi_{\{\cos(\theta-\bar\theta^R_f)<0\}} \nonumber \\
&+ \Lambda(\sigma_f^{R,+},s')\chi_{\{\cos(\theta-\bar\theta^R_f)>0\}}\Big)f(x,\theta+\pi, s',t)\dd s'. 
\label{eq:reverse}
\end{align}
The integral reflects the fact that reversing particles of all ages that point in direction $\theta+\pi$, will add mass to the distribution function at angle $\theta$ and age $s=0$, since reversing resets the biochemical age to zero. An alternative formulation - which we discard here - would be to add a Dirac delta at $s=0$ multiplied by the right-hand side of \eqref{eq:reverse} (divided by $T$) to the expression of \eqref{eqn:QRrev}. Both approaches are equivalent. In particular, the correctness of the mass balance can be checked by integration of \eqref{eq:FP}  with respect to $s$ on $[0,\infty)$.\newline\par
\noindent
\paragraph{Scaling.} Analogous to \cite{Degond2016}, we perform the nondimensionalization and the hydrodynamic scaling in one step. 
On the microscopic scale the reference time and space units are given by $t_0 = 1/\nu$ and $x_0 = v_0 t_0$. 
The age variable $s$ remains unchanged since it is already dimensionless. The scaled diffusion constant is $d = D\,t_0$.
On the macroscopic scale we use the coarse units $t_0' = t_0/\e, x_0' = x_0/\e$ where $\e>0$ is some small real number. Then the dimensionless macroscopic variables are
$\hat t = \frac{t}{t_0'}, \hat x = \frac{x}{x_0'}$. 
Further we set $\hat R = \frac{R}{x_0'}$. The scaled distribution function $\hat f$ and densities $\hat\sigma_f^{\hat R,\pm}$ are given by
\begin{align*}
&\hat f( \hat x, \theta, s, \hat t) = \frac{f(x,\theta, s, t)}{(1/x_0')^2}\quad \text{and}\quad \hat \sigma^{\hat R,\pm}_{\hat f}( \hat x, \hat t) = \frac{\sigma^{R,\pm}_f(x, t)}{(1/x_0')^2}.
\end{align*}
For the reversal term, we set $T = \hat T t_0'$ and $\hat\Lambda(\hat\sigma_{\hat f}^{\hat R,\pm},s) = \Lambda(\sigma^{R,\pm}_f,s) t_0'$. 
Note that $\bar\theta^{\hat R}_{\hat f}(\hat x,\hat t)=\bar\theta^{R}_f(x,t)$. 
%We only outline the next steps, for more details see \cite{Degond2016}. 
The scaled equation reads
\begin{align*}
\partial_{\hat t}\hat f + \nabla_{\hat x}\cdot(v(\theta)\hat f) &+ \frac{1}{\hat T}\partial_s \hat f=\\
&\frac{1}{\e}\partial_\theta\left[{\rm Sign}\left(\cos(\theta-\bar\theta^{\hat R}_{\hat f})\right)\sin(\theta-\bar\theta^{\hat R}_{\hat f})\hat f+d\cos^2(\theta-\bar\theta^{\hat R}_{\hat f})\pd_\theta \hat f\right]\\
& - \left[\Lambda(\hat \sigma_{\hat f}^{\hat R,-},s)\chi_{\{\cos(\theta-\bar\theta^{\hat R}_{\hat f})>0\}} + \Lambda(\sigma_{\hat f}^{\hat R,+},s)\chi_{\{\cos(\theta-\bar\theta^{\hat R}_{\hat f})<0\}}\right] \hat f,
\end{align*}
where the second and third line represent the scaled alignment and reversal operators respectively. The scaled boundary condition at $s=0$ is given by
\begin{align*}
\hat f(\hat x,\theta, 0,\hat t)=\hat T\!\!\int_0^\infty\!\!\Big(&\Lambda(\hat \sigma_{\hat f}^{\hat R,-},s')\chi_{\{\cos(\theta-\bar\theta^{\hat R}_{\hat f})<0\}} \\
&+ \Lambda(\hat \sigma_{\hat f}^{\hat R,+},s')\chi_{\{\cos(\theta-\bar\theta^{\hat R}_{\hat f})>0\}}\Big)\hat f(\hat x,\theta+\pi, s',\hat t)\dd s'.
\end{align*}
At this point the definitions of the nematic mean direction and the densities $\hat \sigma_{\hat f}^{\hat R, \pm}$ still involve space integrals, i.e. they are non-local. We assume purely local interactions for both alignment and reversals and therefore set $\hat R=\e r$ with $r = \mathcal{O}(1)$. 
Then Taylor expansion of $\bar\theta^{\e r}_{\hat f}$ and $\hat \sigma_{\hat f}^{\e r,\pm}$ around $\e=0$ shows that the functions can be approximated by the local-in-space functions $\bar\theta_{\hat f}$ and $\hat \sigma_{\hat f}^{\pm}$ respectively (see \eqref{eqn:thetaBar_local}, \eqref{eqn:sigma_local}) with a remainder of order $\mathcal{O}(\e^2)$. 
%The resulting equation is presented in the next paragraph. 
In the following we drop the hats for better readability and call the solution $f^\e$ to emphasize its dependence on $\e$.\newline\par
\noindent
\paragraph{Hyrodynamic Limit.} To derive the mean field equation, we need to find the solution $f^\e(x,\theta,s,t)$ as $\e \to 0$ in
\begin{align}
\label{eqn:kinetic}
\e\big(\partial_t f^\e + \nabla_x\cdot(v(\theta)f^\e) + \frac{1}{T}\partial_s f^\e\big) 
= Q_\text{al}(f^\e) + \e Q_\text{rev}(f^\e),
\end{align}
where 
\begin{align}
\label{eqn:deriv_Q_al}
&Q_\text{al}(f) = \partial_\theta\left[{\rm Sign}\left(\cos(\theta-\bar\theta_f)\right)\sin(\theta-\bar\theta_f)f+d\cos^2(\theta-\bar\theta_f)\pd_\theta f\right], 
\end{align}
and 
\begin{align*}
Q_\text{rev}(f)=&- \left(\Lambda(\sigma_f^-(x),s)\chi_{\{\cos(\theta-\bar\theta_f)>0\}} + \Lambda(\sigma_f^+(x),s)\chi_{\{\cos(\theta-\bar\theta_f)<0\}}\right)f(x,\theta,s,t),
\end{align*}
supplemented by the boundary condition at $s=0$
\begin{align*}
f(x,\theta,0,t)=&T\int_0^\infty\!\!\!\Big(\Lambda(\sigma_f^-(x),s')\chi_{\{\cos(\theta-\bar\theta_f)<0\}} \\& +\Lambda(\sigma_f^+(x),s')\chi_{\{\cos(\theta-\bar\theta_f)>0\}}\Big)
    f(x,\theta+\pi, s',t)\dd s'.
\end{align*}
The mean nematic direction is defined by 
\begin{align}
\label{eqn:thetaBar_local}
v(2\bar\theta_f(x, t)) = \frac{J_f(x,t)}{|J_f(x,t)|}, \quad \text{with } 
J_f(x,t) = \int_0^\infty\int_{-\pi}^{\pi}v(2\theta)f(x, \theta, s, t)\dd\theta \dd s, 
\end{align}
and the local mass functions are  
\begin{align}
\label{eqn:sigma_local}
\sigma_{f}^{\pm} (x,t) = \int_0^\infty\int_{\pm\cos(\theta - \bar\theta_f)>0} f(x,\theta,s,t)\dd\theta \dd s.
\end{align}
Taking the hydrodynamic limit in the SOH framework \cite{Degond2008} now involves two steps: (i) Characterizing the kernel of $Q_\text{al}(f)$ ($\theta$ dependence) and (ii) Using Generalized Collision Invariances (GCIs) to extract information about the $x$ and $t$ dependence from the transport and reversal terms.\newline
Step (i): Since  $Q_\text{al}(f)$ is similar to the collision operator analyzed in \cite{Degond2016}, but for the addition of the age variable, we simply state the result without proof:
\begin{lemma}
The kernel of $Q_\text{al}$ is given by
\begin{align*}
\left\{\bar f_{\rho_+(s),\rho_-(s),\bar\theta}(\theta)|\rho_\pm: [0,\infty)\rightarrow [0,\infty), \bar\theta\in[0,\pi) \right\},
\end{align*}
where
\begin{align}
\label{eqn:equilibria}
&\bar f_{\rho_+,\rho_-,\bar\theta}(\theta)=
\begin{cases}
\rho_+M_{\bar\theta}(\theta)\quad \text{for}\quad \cos(\theta-\bar\theta)>0\\
\rho_-M_{\bar\theta}(\theta)\quad \text{for}\quad \cos(\theta-\bar\theta)<0.
\end{cases}
\end{align}
$M_{\bar\theta}(\theta)$ describes the Generalized von Mises (GVM) distribution defined by
\begin{align*}
&M_{\bar\theta}(\theta)=\frac{1}{Z_d}\exp{\left(-\frac{1}{d|\cos{(\theta-\bar\theta)}|}\right)}, \quad \theta\in[-\pi,\pi)\\
&\text{where}\quad Z_d=\int_{\cos{\theta}>0}\exp{\left(-\frac{1}{d\cos\theta}\right)}\dd \theta.
\end{align*} 
\end{lemma}
\begin{figure}[t!]
\centering
\includegraphics[width=0.75\textwidth]{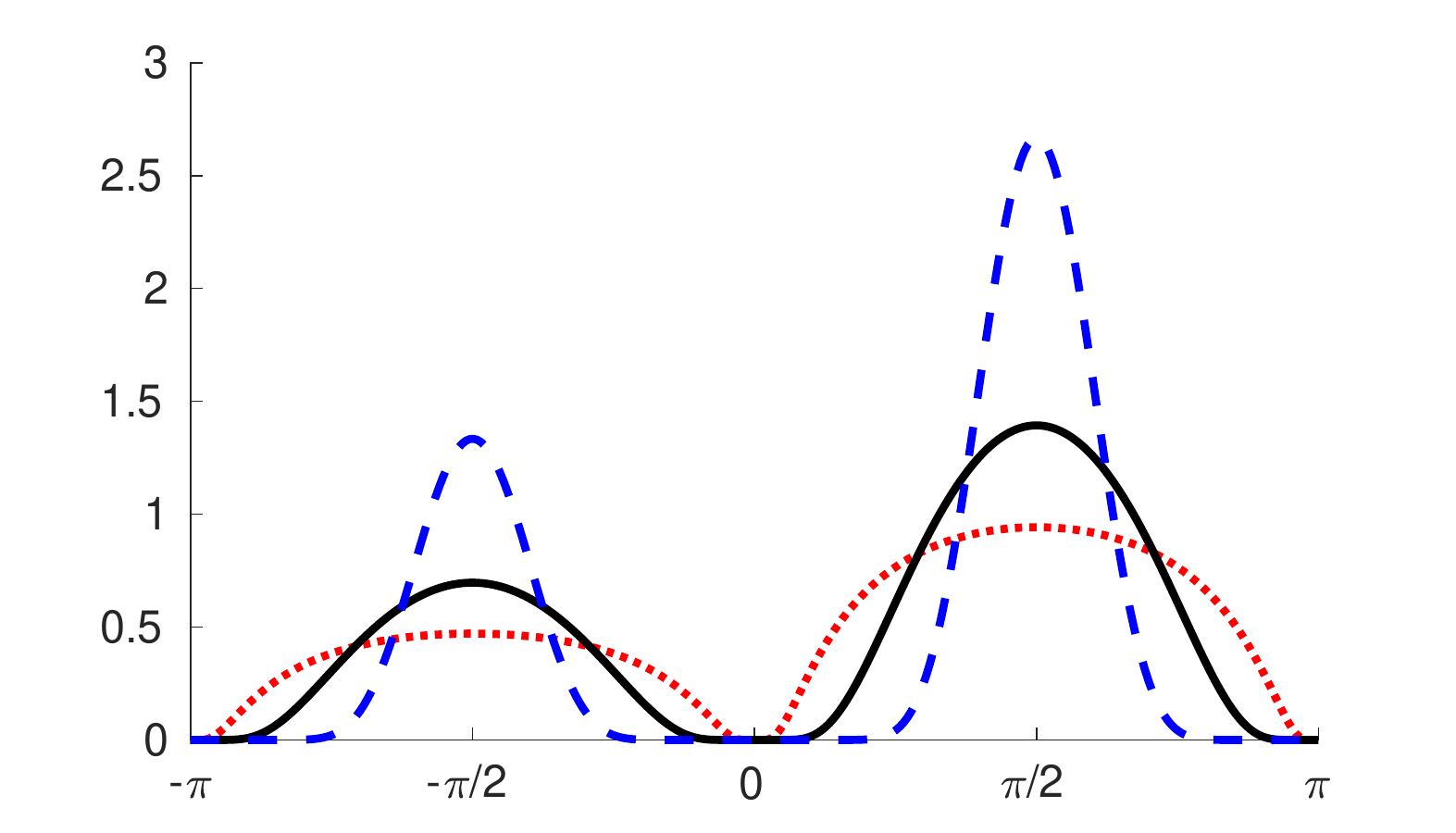}
\caption{Equilibria of $Q_\text{al}$. The equilibrium distribution $\bar f$ as defined in \eqref{eqn:equilibria} for $(\rho_+,\rho_-, \bar\theta)=(2,1,\pi/2)$ using $d=2, 0.5$ and $0.1$ (red-dotted, black-solid and blue-dashed respectively).}
\label{fig:equilibria}
\end{figure}
The equilibria have the shape of two opposing maxima: One in direction $\bar\theta$ with mass $\rho_+(s)$ and one in direction $\bar\theta-\pi$ with mass $\rho_-(s)$ and are depicted in Fig.\,\ref{fig:equilibria}.\newline
Step (ii): In a classical setting, one would at this point mutliply equation \eqref{eqn:kinetic} with collision invariants (CIs) and integrate over all directions $\theta$ and all ages $s$. A CI is defined as a function $\Psi(\theta,s)$ on $[-\pi,\pi)\times [0,\infty)$, such that
\[
\iint Q_\text{al}(f)\Psi(\theta,s)\dd\theta\dd s=0 \qquad \forall f.
\]
This would allow us to remove the term of order one from \eqref{eqn:kinetic} and thereby yield three macroscopic equations that describe the constant $\bar\theta$ and the two functions $\rho_\pm(s)$ that characterize the equilibria defined in \eqref{eqn:equilibria}. However, the operator $Q_\text{al}$ has only one family of CIs: $\Psi(\theta, s)\equiv \psi_0(s)$, for an arbitrary function $\psi_0$ which represents mass conservation of particles of a given age $s$ (since alignment doesn't modify the age). This necessitates the use of Generalized Collision Invariants (GCIs). Given an angle of lines $\bar\theta$ a GCI associated with $\bar\theta$ is a function $\Psi_{\bar\theta}(\theta,s)$ defined on $[-\pi,\pi)\times[0,\infty)$, such that
\[
\iint Q_{\text{al},\bar\theta}(f)\Psi_{\bar\theta}(\theta,s)\dd\theta\dd s=0 \qquad \forall f \quad \text{with} \quad \bar\theta_f=\bar\theta,
\]
where $Q_{\text{al},\bar\theta}$ is defined analogously to $Q_\text{al}$ in \eqref{eqn:deriv_Q_al}, but with $\bar\theta_f$ replaced by $\bar\theta$.
A GCI is a more general concept than CIs. If enough families of GCIs can be found, Step (ii) of the derivation now requires integrating equation \eqref{eqn:kinetic} against these GCIs associated to $\bar\theta_{f^\e}$. Similar to the classical CI-based approach, this removes the leading order singular term and allows us to derive the macroscopic equations. We omit the precise functional analytical setting as it can be easily extended from \cite{Degond2016} and simply state the following result:
\begin{lemma}
Given an angle of lines $\bar\theta$, the space of GCIs of $Q_\text{al}$ associated with $\bar\theta$ is spanned by $\psi_{\bar \theta}^\pm(\theta,s):=\text{H}(\pm \cos{(\theta-\bar\theta)})\psi_0(s)$ and $g_{\bar\theta}(\theta,s):=g(\theta-\bar\theta)$ where $\text{H}$ denotes the Heaviside function, 
\[
g(\theta)=-\int_0^\theta \frac{\int_\beta^{\pi/2}\sin{2\alpha}\exp{\left(-\frac{1}{d\cos\alpha}\right)}\dd\alpha}{\cos^2\beta\exp{\left(\frac{1}{d\cos\beta}\right)}}\dd\beta \quad \text{for}\quad \theta\in[0,\pi/2]
\]
which is extended to $[-\pi,\pi)$ by $g(-\theta)=-g(\theta)$ and $g(\pi-\theta)=-g(\theta)$, and $\psi_0(s)$ is an arbitrary functions of $s$.
\end{lemma}
Note that $\psi_{\bar \theta}^+(\theta,s)+\psi_{\bar \theta}^-(\theta,s)\equiv \psi_0(s)$ and we recover the CI associated with mass conservation.
Now we proceed as explained, by integrating \eqref{eqn:kinetic} against the three GCIs associated to $\bar\theta_{f^\e}$. \newline\par
\noindent
Proceeding similar to \cite{Degond2016} finally yields
\begin{proposition}
\label{thm:limit_cont_agestructured}
Taking the (formal) limit $\e\rightarrow 0$ in \eqref{eqn:kinetic} we obtain
\begin{align*}
f^\e(x,\theta,s,t)\, \, \longrightarrow \, \, \bar f_{\rho_+(x,s,t), \rho_-(x,s,t), \bar \theta(x,t)}(\theta), 
\end{align*}
where $\bar f_{\rho_+(x,s,t), \rho_-(x,s,t), \bar \theta(x,t)}(\theta)$ is given by \eqref{eqn:equilibria} and the macroscopic quantities $\rho_\pm(x,s,t)$ and $\bar\theta(x,t)$ are such that $\rho_\pm(x,s,t)\in[0,\infty)$ and $\bar\theta(x,t)$ is a real number defined modulo $\pi$, and fulfill
\begin{subequations}
\label{eqn:SOH_age_scaled}
\begin{align}
& \pd_t \rho_+ + d_1 \nabla_x\cdot (\rho_+ v(\bar \theta))+\frac{1}{T}\pd_s\rho_+=-\Lambda(\sigma_-, s)\rho_+,\\
& \pd_t \rho_- - d_1 \nabla_x\cdot (\rho_- v(\bar \theta))+\frac{1}{T}\pd_s\rho_-=-\Lambda(\sigma_+, s)\rho_-,\\
&(\sigma_++\sigma_-)\pd_t \bar\theta+d_2(\sigma_+-\sigma_-)(v(\bar\theta)\cdot \nabla_x)\bar\theta+\mu\,v(\bar\theta)^\perp\cdot\nabla_x(\sigma_+-\sigma_-)=0,
\end{align}
\end{subequations}
supplemented by the boundary conditions
\begin{align}
\label{eqn:SOH_age_scaled_bc}
\rho_+(x,0,t)= T \int_0^\infty \Lambda(\sigma_+,s)\rho_- \dd s, \quad
\rho_-(x,0,t)= T \int_0^\infty \Lambda(\sigma_-,s)\rho_+ \dd s,
\end{align}
where the coefficients $d_1$, $d_2$ and $\mu$ are given by
\begin{align}\label{SOH_para}
d_1 = \langle \cos \rangle_M, \qquad d_2 = \frac{\langle g\frac{\sin}{\cos}\rangle_M}{\langle g\frac{\sin}{\cos^2}\rangle_M},
\qquad \mu = d\frac{\langle g\sin\rangle_M}{\langle g\frac{\sin}{\cos^2}\rangle_M},
\end{align}
$v(\bar\theta)^\perp=(-\sin(\bar\theta), \cos(\bar\theta))^T$ and $\langle \phi \rangle_M$ represents the average with respect to $M(\theta) = M_0(\theta)$:
\[
\langle \phi \rangle_M = 2\int_0^{\frac{\pi}{2}}\phi(\theta)M(\theta)\dd\theta
= \frac{\int_0^{\frac{\pi}{2}}\phi(\theta) e^{-\frac{1}{d\cos\theta}}\dd\theta}{\int_0^{\frac{\pi}{2}}e^{-\frac{1}{d\cos\theta}}\dd\theta}.
\]
\end{proposition}
Finally we remove the non-dimensionalization and revert system \eqref{eqn:SOH_age_scaled}-\eqref{eqn:SOH_age_scaled_bc} back to physical units yielding \eqref{eqn:macro_main}-\eqref{eqn:macro_main_bc} (below). The next section summarizes the model.

\subsubsection{The macroscopic continuous-age model.} We denote the spatial variable by $x\in\mathbb{R}^2$, the age variable by $s\in[0,\infty)$ and time by $t>0$. Then $\bar\theta(x,t)\in[0,\pi)$ describes the local nematic mean direction, which is independent of $s$. We recall that $v(\bar\theta):=\left(\cos(\bar\theta), \sin(\bar\theta)\right)^T$ and define $v^\perp$ as its left-oriented orthogonal.  We denote by $\rho_{+}(x,s,t)$ and $\rho_{-}(x,s,t)$ the local densities of bacteria of age $s$ that are transported in the directions $v(\bar\theta)$ and $-v(\bar\theta)$ respectively. $\bar\theta(x,t)$, $\rho_\pm(x,s,t)$ fulfill the following system of equations:
\begin{subequations}\label{eqn:macro_main}
\begin{align}
& \pd_t \rho_+ + d_1 v_0\nabla_x\cdot (\rho_+ v(\bar \theta))+\frac{1}{T}\pd_s\rho_+=-\Lambda(\sigma_-, s)\rho_+,\\
& \pd_t \rho_- - d_1 v_0\nabla_x\cdot (\rho_- v(\bar \theta))+\frac{1}{T}\pd_s\rho_-=-\Lambda(\sigma_+, s)\rho_-,\\
&(\sigma_++\sigma_-)\pd_t \bar\theta+d_2 v_0(\sigma_+-\sigma_-)(v(\bar\theta)\cdot \nabla_x)\bar\theta+\mu v_0 v(\bar\theta)^\perp\cdot\nabla_x(\sigma_+-\sigma_-)=0,
\end{align}
\end{subequations}
supplemented by boundary conditions at the age $s=0$,
\begin{align}
\label{eqn:macro_main_bc}
\rho_+(x,0,t)=T \int_0^\infty \Lambda(\sigma_+,s)\rho_- \dd s, \qquad
\rho_-(x,0,t)=T \int_0^\infty \Lambda(\sigma_-,s)\rho_+ \dd s,
\end{align}
where the coefficients $d_1$, $d_2$ and $\mu$ are given by \eqref{SOH_para}, $\Lambda(\sigma,s)$ is defined in \eqref{eqn:Lambda} and $\sigma_\pm(x,t)$ are the local masses of the two opposing groups:
\[
\sigma_\pm(x,t) = \int_0^\infty \rho_\pm(x,s,t)\dd s.
\]

\begin{figure}[t!]
\centering
\subfigure[Continuous-age model]{\includegraphics[width=0.4\textwidth]{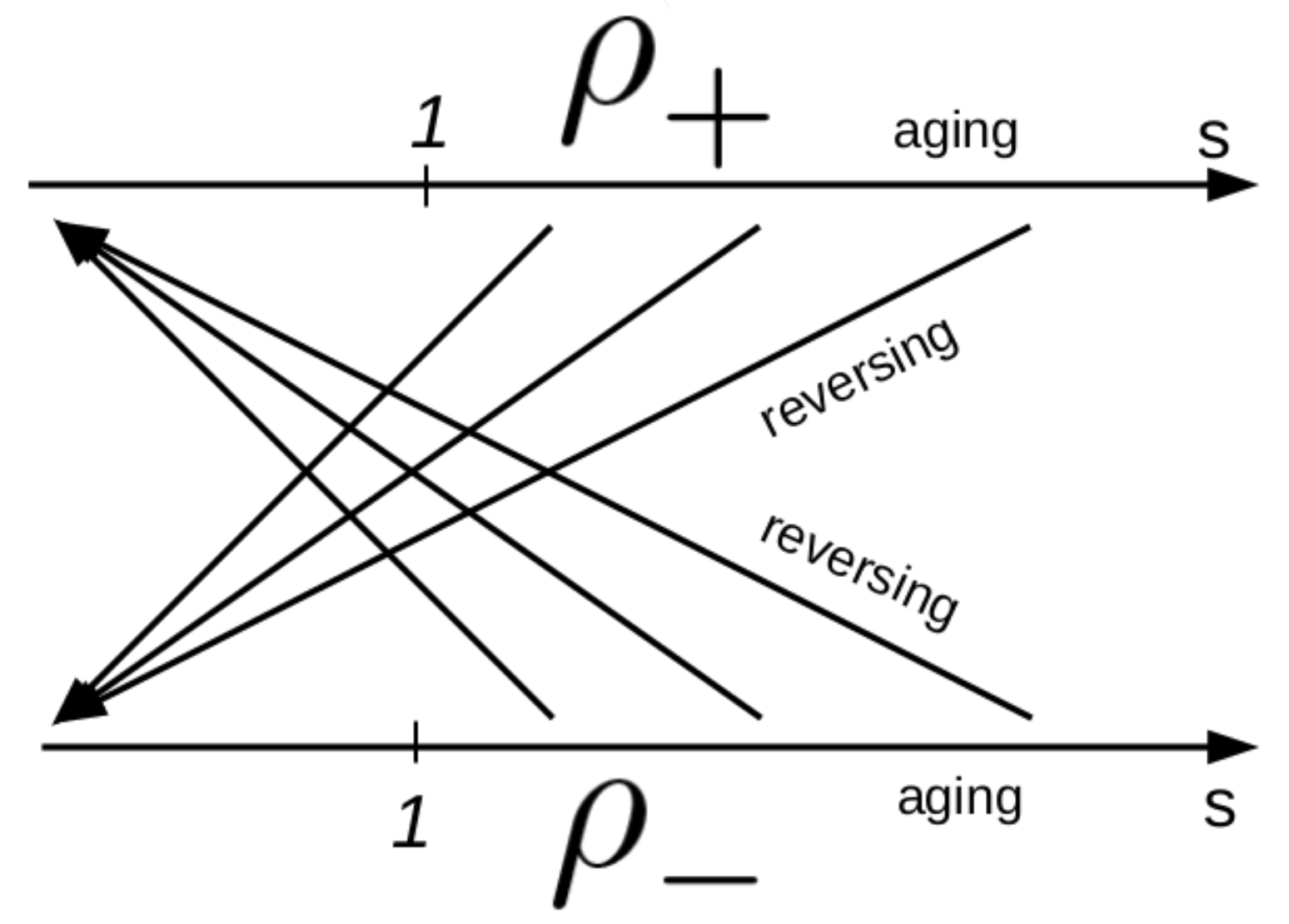}}
\subfigure[Two-age model]{\includegraphics[width=0.28\textwidth]{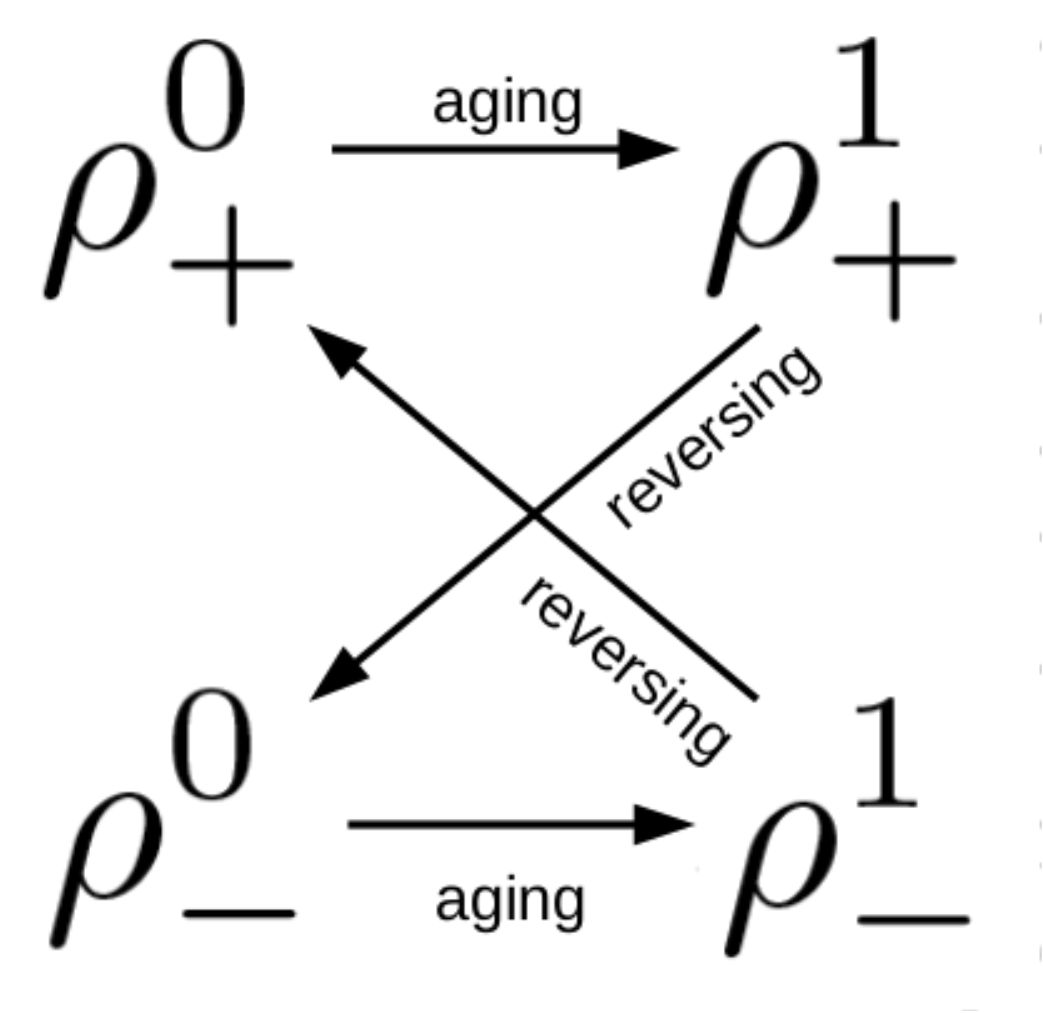}}
\subfigure[Memory-free model]{\includegraphics[width=0.24\textwidth]{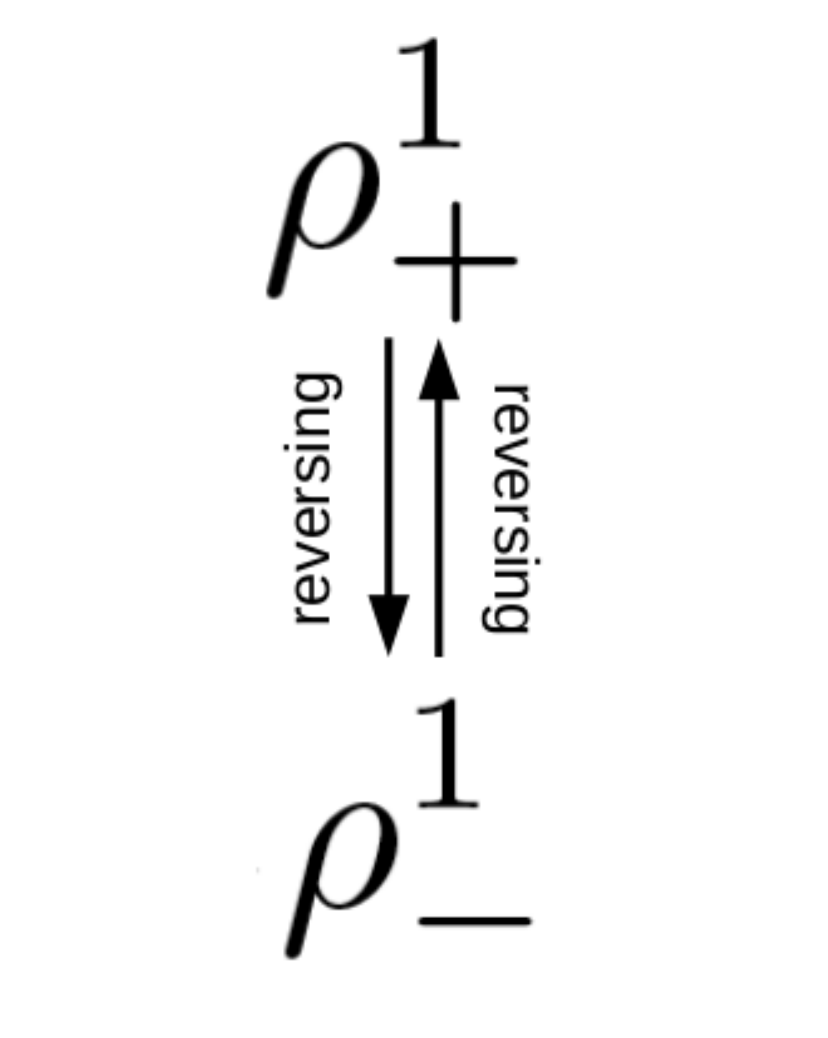}}
\caption{Reaction terms in the three macroscopic models. In the continuous-age model, (a) the densities $\rho_\pm$ depend on the continuous age variable $s$. Particles can \emph{age} along $s$ (horizontal arrows) or reverse, if $s>1$ (diagonal arrows), which lets them join the other group at age $s=0$. In the 2-age model (b), there are only 2 age groups, insensitive to C-signaling or sensitive to C-signaling, denoted by the superscripts $0$ and $1$. In the memory-free model (c), all particles are sensitive to C-signaling and there is no aging. }
\label{fig:reaction_schematic}
\end{figure}

Fig.\,\ref{fig:reaction_schematic}(a) illustrates the reversal- and age-related dynamics of \eqref{eqn:macro_main}-\eqref{eqn:macro_main_bc}: For $s \in [0,1]$, bacteria are in the refractory period and can not reverse, ensured by $\Lambda(\sigma_\mp, s) = 0$.
For $s > 1$, bacteria enter the C-factor sensitive period and the reversal frequency is governed by the reversal function $\Lambda(\sigma_\mp, s) = \lambda(\sigma_\mp)$ (using \eqref{eq:phi} and \eqref{eqn:Lambda}). Assuming that there are in fact only two essential age states for the age variable $s$ allows to simplify \eqref{eqn:macro_main} to remove the age $s$ as independent variable. This is done in the following section.

%%%%
% 2-AGE MODEL
%%%%

\subsection{The 2-Age Model}
\label{ssec:cont_2age}

To arrive at an easy-to-handle, yet powerful macroscopic model, we perform one last simplification step: we assume only two age group of bacteria, those who can, and those who cannot reverse. The main difference to the continuous-age model is that the aging itself is now described as a simple reaction term with rate $1/T$. Fig.\,\ref{fig:reaction_schematic}(b) depicts the corresponding reaction schematic. The time evolution of the nematic mean direction remains unchanged. The derivation is given below in Sec.\,\ref{ssec:derivation_2age}.

%%%%%%%%%
%  DERIVATION II: 2-AGE
%%%%%%%%%

\subsubsection{Derivation of the 2-age model}
\label{ssec:derivation_2age}

\paragraph{An age-discretized macro-model with $K+1$ age groups.}
As a starting point we use the system of equations \eqref{eqn:macro_main} together with the boundary conditions \eqref{eqn:macro_main_bc}, i.e. the macroscopic myxobacteria model with a continuous age variable. 
To derive the corresponding discrete age system we discretize the age variable $s$ by $s_k=k\Delta s$ for $k=0, \ldots, K$, yielding $K+1$ age groups defined by
\[
\tilde\rho_\pm^k(x,t):=\rho_\pm(x,s_k,t), \quad k=0, \ldots ,K.
\]
To get equations for $\tilde\rho_\pm^k(x,t)$ we use a forward difference discretization of $\partial_s \rho_\pm$. Since the equation for $\bar\theta$ is independent of $s$, we only need to consider the two density equations. The only noteworthy point is that the largest age group i.e. $\tilde\rho^{K}_\pm$ can lose particles only by reversing, not by aging. 
The corresponding system for a discrete age system with $K+1$ age groups is
\begin{align}
\label{eqn:macro_Np1ages}
\partial_t\tilde\rho_+^k+d_1v_0\nabla_x\cdot (\tilde\rho_+^k v(\bar\theta))&+\frac{1}{T\Delta s}\left(\tilde\rho_+^k-\tilde\rho_+^{k-1}\right)\nonumber\\
&=-\lambda(\tilde\sigma_-)\phi(s_k)\tilde\rho_+^k,
\quad \text{ for } k=0,...,K-1, \nonumber\\
\partial_t\tilde\rho_+^{K}+d_1v_0\nabla_x\cdot (\tilde\rho_+^{K} v(\bar\theta))&-\frac{1}{T\Delta s}\tilde\rho_+^{K-1}=-\lambda(\tilde\sigma_-)\phi(s_{K})\tilde\rho_+^{K},
\quad \text{ for } k=K
\end{align}
(and analogously for $\tilde\rho_-^k$). The boundary conditions are included by using ``virtual" age groups defined by discretizing the integrals in \eqref{eqn:macro_main_bc}, yielding
\begin{align*}
\tilde\rho_+^{-1}:=T\Delta s\lambda(\tilde\sigma_+)\sum_{k=0}^{K}\tilde\rho_-^k\phi(s_k), \quad \tilde\rho_-^{-1}:=T\Delta s\lambda(\tilde\sigma_-)\sum_{k=0}^{K}\tilde\rho_+^k\phi(s_k).
\end{align*}
The total group densities $\tilde\sigma_\pm(x,t)$ are defined by
\begin{align*}
\tilde\sigma_+=\Delta s\sum_{k=0}^{K}\tilde\rho_+^k, \quad \tilde\sigma_-=\Delta s\sum_{k=0}^{K}\tilde\rho_-^k.
\end{align*}
\noindent
\paragraph{The 2-age macro-model.} To obtain the 2-age model \eqref{eqn:macro_2age}, we define
\begin{align*}
\rho_\pm^0(x,t):=\Delta s\,\tilde\rho_\pm^0(x,t),\qquad \rho_\pm^1(x,t):=\Delta s\sum_{k=1}^{K}\tilde\rho_\pm^k(x,t)
\end{align*}
in \eqref{eqn:macro_Np1ages} and assume that $\phi(s_0)=0$ and $\phi(s_k)=1$ for $k\geq 1$. This yields a closed system for $(\rho^0_\pm, \rho^1_\pm)$. For better readability we drop the tilde on the expressions for the total densities $\sigma_\pm$.
\begin{align*}
& \pd_t \rho^0_+ + d_1 v_0\nabla_x\cdot (\rho^0_+ v(\bar \theta))=-\frac{1}{T\Delta s}\rho^0_++\lambda(\sigma_+)\rho_-^1,\\
& \pd_t \rho^1_+ + d_1 v_0\nabla_x\cdot (\rho^1_+ v(\bar \theta))=\frac{1}{T\Delta s}\rho^0_+-\lambda(\sigma_-)\rho_+^1,\\
& \pd_t \rho^0_- - d_1 v_0 \nabla_x\cdot (\rho^0_- v(\bar \theta))=-\frac{1}{T\Delta s}\rho^0_-+\lambda(\sigma_-)\rho_+^1,\\
& \pd_t \rho^1_- - d_1 v_0\nabla_x\cdot (\rho^1_- v(\bar \theta))=\frac{1}{T\Delta s}\rho^0_--\lambda(\sigma_+)\rho_-^1.
\end{align*}
Setting $\Delta s=1$ we get the final system \eqref{eqn:macro_2age_diffFree}, below. In Sec.\,\ref{sec:numerics} we demonstrate that this 2-age model is a good approximation of the full age-dependent dynamics and sufficient to reproduce and explain almost all experimentally observed features of myxobacteria.

\subsubsection{The 2-age model}

 The final 2-age model derived in Sec.\,\ref{ssec:derivation_2age} consists of five equations for the four densities $\rho_\pm^{0,1}(x,t)$  and the nematic mean direction $\bar\theta(x,t)$. Subindices $\pm$ denote densities of bacteria moving with and against the nematic mean direction respectively and superindices $0/1$ denote bacteria in a non-reversible and reversible state respectively.
\begin{subequations}
\label{eqn:macro_2age_diffFree}
\begin{align}
& \pd_t \rho^0_+ + d_1 v_0\nabla_x\cdot (\rho^0_+ v(\bar \theta))=-\frac{1}{T}\rho^0_++\lambda(\sigma_+)\rho_-^1,\\
& \pd_t \rho^1_+ + d_1 v_0\nabla_x\cdot (\rho^1_+ v(\bar \theta))=\frac{1}{T}\rho^0_+-\lambda(\sigma_-)\rho_+^1,\\
& \pd_t \rho^0_- - d_1 v_0 \nabla_x\cdot (\rho^0_- v(\bar \theta))=-\frac{1}{T}\rho^0_-+\lambda(\sigma_-)\rho_+^1,\\
& \pd_t \rho^1_- - d_1 v_0\nabla_x\cdot (\rho^1_- v(\bar \theta))=\frac{1}{T}\rho^0_--\lambda(\sigma_+)\rho_-^1,
\end{align}
\end{subequations}
\begin{align}
\label{eqn:macro_2age_theta}
&(\sigma_++\sigma_-)\pd_t \bar\theta+d_2v_0(\sigma_+-\sigma_-)(v(\bar\theta)\cdot \nabla_x)\bar\theta+\mu v_0\,v(\bar\theta)^\perp\nabla_x(\sigma_+-\sigma_-)=0,
\end{align}
where the local masses are given by $\sigma_\pm=\rho_\pm^0+\rho_\pm^1$. The constants $d_1$, $d_2$ and $\mu$ are defined analogously as for \eqref{eqn:macro_main}. Note that this model does not contain any spatial diffusion, which we will discuss further in Sec.\,\ref{ssec:add_diffusion}.

%%%%
% LAMBDA
%%%%

\subsection{The Reversal Frequency $\lambda(\rho)$} \label{ssec:reversal}
To complete the models presented in Secs.\,\ref{ssec:particle}-\ref{ssec:cont_2age}, the density dependence of the reversal frequency $\lambda(\rho)$ has to be specified based on the available information from experiments.
Firstly, in most experiments the number of reversals increase with the density of opposing bacteria, i.e. $\lambda'(\rho)\geq 0$. 
Next, in the absence of other bacteria isolated myxobacteria still reverse, i.e. $\lambda(0)=:\lambda_m>0$. Spontaneous reversal rates between $0.07$ and $0.09$ reversals per minute have been reported \cite{Sager1994,Sliusarenko2006,Welch2001}. 
Thirdly, there seems to be an upper limit as to how frequent reversals can be, which confirms the biological intuition that the rearrangement of the internal movement machinery takes some time. 
In \cite{Welch2001} a maximal rate of $1.5$ reversals per minute have been observed. 
Finally, in \cite{Sager1994} experiments were performed in which the reversal rate of isolated bacteria was measured in response to externally adding the signaling molecule C-factor, which was thought to communicate the density information. 
At low concentration of C-factor the reversal rate remained the same, while with increasing concentration a growth in reversal rate was observed, which plateaued for very high concentrations of C-factor. Using these pieces of information we assume a sigmoid shape of $\lambda(\rho)$. 
As a convenient representation we use a $C^1$ and piecewise smooth function:
\begin{align}
\label{eqn:lambda}
\lambda(\rho)=\begin{cases}
\lambda_m+\frac{1}{2}(\lambda_M-\lambda_m)(\frac{\rho}{\bar\rho})^2 &\text{if }\rho <\bar\rho,\\
\lambda_M-\frac{1}{2}(\lambda_M-\lambda_m)(\frac{\rho}{\bar\rho}-2)^2 &\text{if }\bar\rho\leq \rho <2\bar\rho,\\
\lambda_M &\text{elsewhere}.
\end{cases}
\end{align}
Note that $\lambda(\rho)$ is parameterized by three quantities, the spontaneous reversal rate $\lambda_m$, the maximal reversal rate $\lambda_M$ and the inflection density $\bar\rho$, at which $\lambda(\rho)$ grows the fastest.

%%%%%%%%%%%%%%%%%%%%%%%%%
%%%%%%%%%%%%%%%%%%%%%%%%%
%%%%
%%%%  	NUMERICAL ANALYSIS
%%%%
%%%%%%%%%%%%%%%%%%%%%%%%%
%%%%%%%%%%%%%%%%%%%%%%%%%

\section{Numerical Analysis - Comparison to Experiments}
\label{sec:numerics}
Depending on the precise experimental set-up, various bacterial speeds $v_0$ were observed \cite{Sliusarenko2006,Sager1994}, ranging from $2.7$ to $11\,\mu m$/min.  We do not have reliable biological data on the refractory period $T$ and the inflection density $\bar\rho$. 
In \cite{Igoshin2001} it was suggested that the refractory period must be less than 40 sec, while in \cite{Borner2002} times around 5 min were suggested. 
We use $T=2$ min, but note that our analysis shows wave synchronization also for any value larger that $\approx 1$ min. For $\bar\rho$ we estimated that it should be of the order of half of the initial mean density to have an effect. A list of all parameters can be found in Tab.\,\ref{tab:parameters}.

%%%%
% IBM 2D
%%%%

\subsection{The particle model in 2D}
\label{ssec:numerics_IBM}

\begin{figure}[t!]
\centering
\subfigure[Ripples in the IBM]{\includegraphics[width=0.342\textwidth]{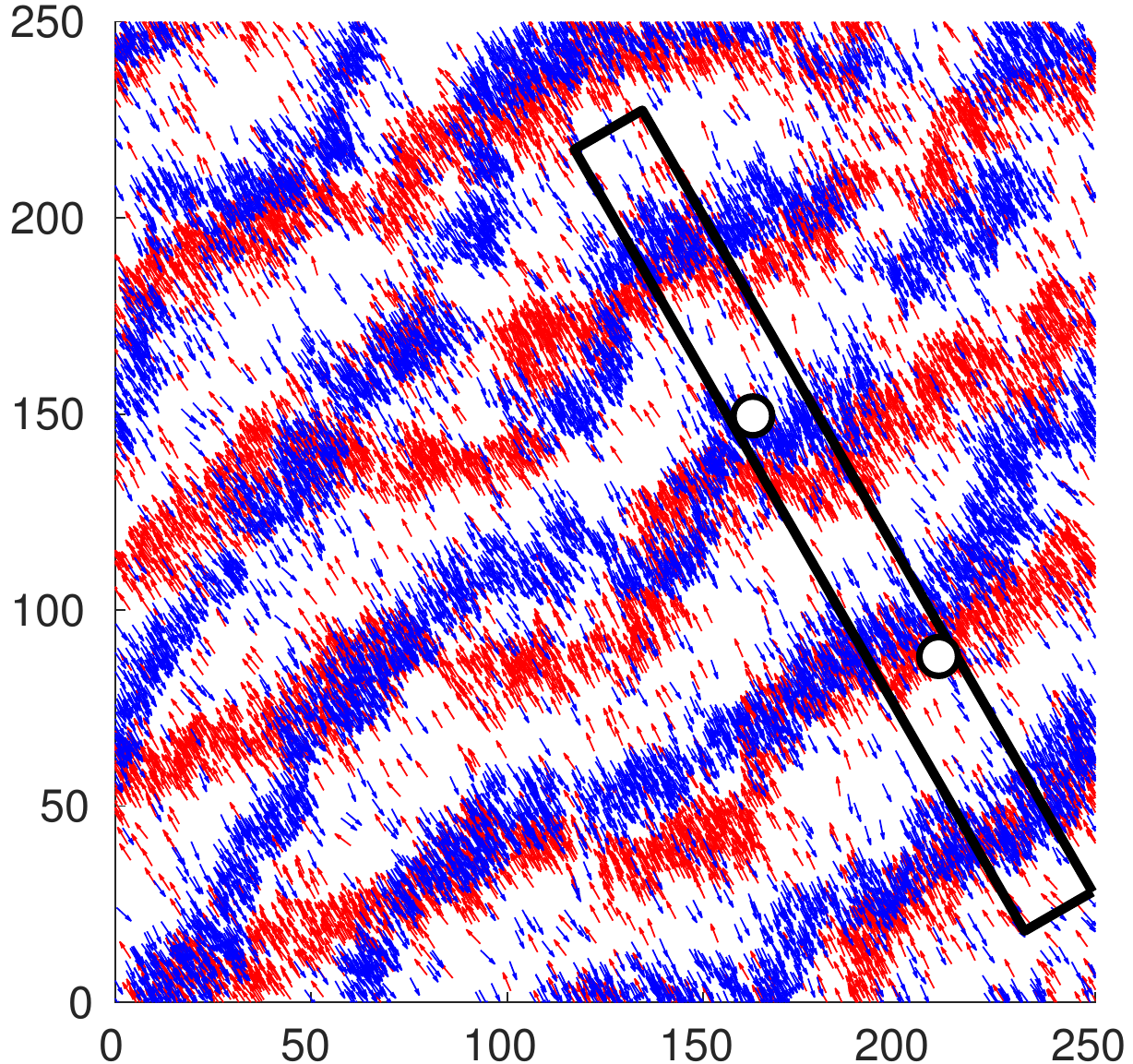}}
\subfigure[Comparison to 2-Age Model]{\includegraphics[width=0.649\textwidth]{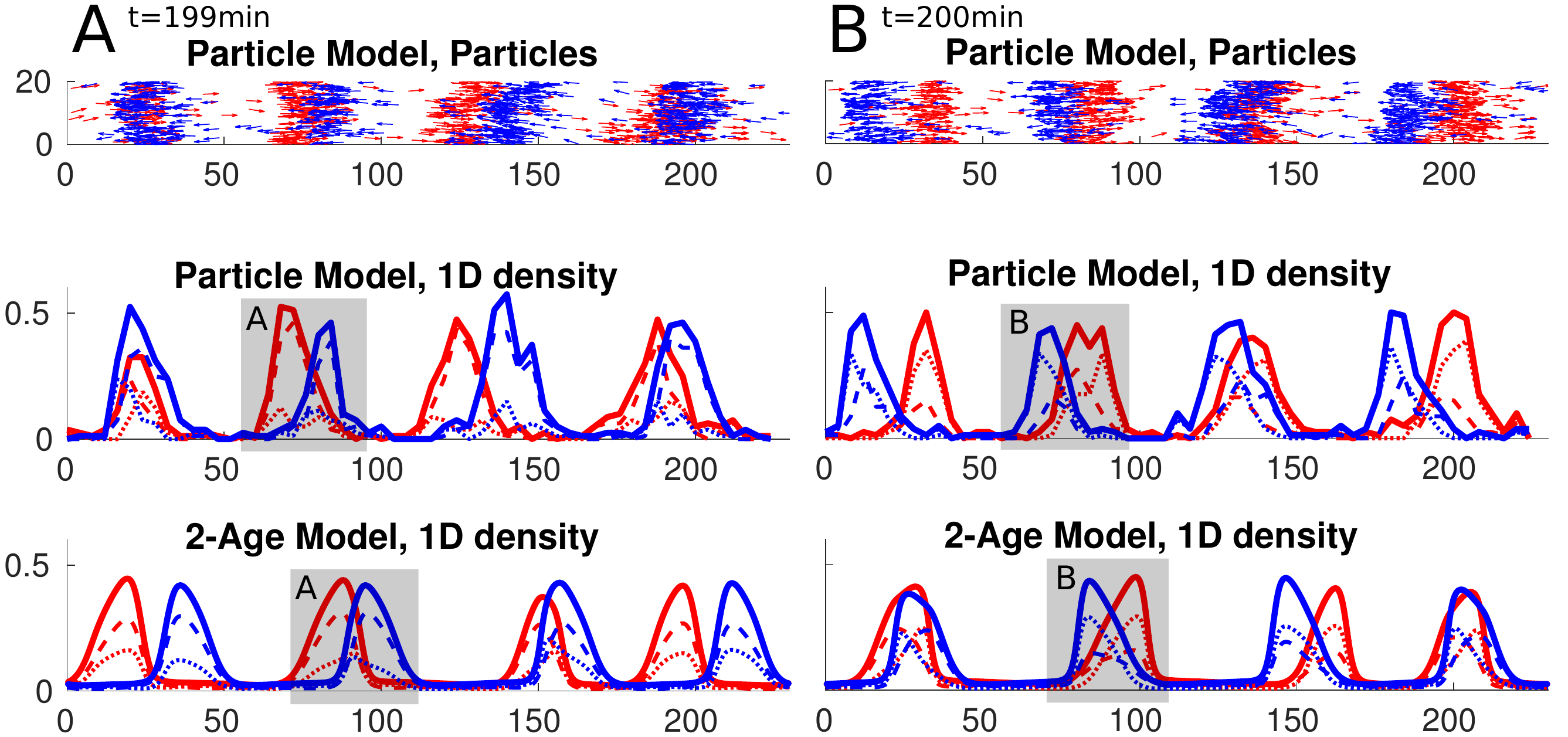}}
\caption{Comparison IBM and 2-Age Model: Wave composition. (a) shows the bacteria at time $t=200$ min of a simulation of the IBM using uniform, random initial conditions (see also SI video 1 for the whole time history of this simulation). The nematic mean direction is almost globally constant with $\bar\Theta=120^\circ$. Bacteria aligned and anti-aligned with it are depicted in red and blue respectively. Units are in $\mu m$. (b) Comparing the IBM to the 2-age model at time $t=199$ min (A) and $t=200$ min (B). Upper row: The rectangular strip marked in (a) turned by the global nematic mean direction. Middle row: The 1D-densities of right-moving (red) and left-moving (blue) bacteria calculated from the strip above with a grid-size of $4\,\mu m$ in $x$ direction and averaging in $y$ direction. Each wave consists of C-signaling insensitive, refractory bacteria ($s_i\leq 1$, dotted) and C-signaling sensitive, non-refractory bacteria ($s_i>1$, dashed). 
Lower row: Simulation of the 1D 2-age model \eqref{eqn:macro_2age} using as initial conditions a uniform random perturbation of magnitude $0.02\,\mu m^{-2}$ of \eqref{eqn:steady}. Color and line-styles are the same as in middle figure. Boxes A and B mark crests before and after collisions respectively, for both the IBM and the 2-age model. y-units for the middle and lower row are in bacteria per $\mu m^2$. All other units are in $\mu m$.}
\label{fig:IBM1}
\end{figure}

As a first test of the model, we simulate the full 2D particle model described in Sec.\,\ref{ssec:particle}. Details about the numerical method as well as simulation parameters can be found in App.\,\ref{app:numerics}.
Both the initial positions $X_i\in\mathbb{R}^2$ and the initial orientations $\Theta_i\in[-\pi,\pi)$ follow a uniform random distribution. The age variable was initialized with a uniform random distribution on $[0,1]$, i.e. all bacteria are assumed to have reversed before the start of the simulation, however also other choices will lead to the same behavior.\newline\par
\noindent
\paragraph{The particle model shows ripple formation.} Within about one hour an almost spatially constant nematic mean direction is established (in this simulation it is $\bar\Theta=120^\circ$) and all bacteria are either aligned or anti-aligned with it, with small deviations caused by the noise. Just like in experiments, macroscopic traveling density bands develop, in which bacteria travel in the same direction as the band itself and normal to its longitudinal axis. Fig.\,\ref{fig:IBM1}(a) depicts the simulation outcome at time $t=200$ min (see also SI video 1 for the whole time history of this simulation). To distinguish between bacteria aligned or anti-aligned with the (global) nematic mean direction, they are shown in red and blue respectively and in the following we will sometimes refer to them as \emph{right- and left-moving} bacteria. The global ordering indicates that the nematic alignment quickly drives the system to a quasi 1D situation. 
To further analyze what happens along the nematic mean direction, we calculate the densities of the right- and left-moving bacteria within a thin strip in the simulation domain (Fig.\,\ref{fig:IBM1}(b) upper and middle rows). 
Additionally we examine the composition of each wave in terms of C-signaling sensitive and insensitive bacteria: the middle rows in Fig.\,\ref{fig:IBM1}(b) clearly show the density waves and indicates that the wave composition is different before (box A) and after (box B) a wave crest collision. This will be examined further below.\newline\par
\noindent
\paragraph{Individual bacteria reverse upon crest collisions.} From biological experiments it is known that when two waves meet, bacteria in the crests typically reverse their direction of movement.  The macroscopic, experimental observations, i.e. that counter-propagating waves move with approximately the bacterial speed, persist over time and travel through each other seemingly unaffected, can be easily verified in Fig.\,\ref{fig:IBM2}, which shows a space-time plot of the total (1D) densities in the rectangular strip marked in Fig.\,\ref{fig:IBM1}(a) for 170 min $\leq t\leq 200$ min. To examine the behavior of individual bacteria in this macroscopic context, Fig.\,\ref{fig:IBM2}(a) also shows the space-time path (blue) of two individual bacteria (marked in Fig.\,\ref{fig:IBM1}(a)). Consistent with experiments, bacteria mostly reverse upon crest collision and rarely in between. This shows that the waves are mostly reflected off each other, confirming the accordion-like behavior known biologically.

%%%%%%%%%%
% ADDING DIFFUSION
%%%%%%%%%%

\subsection{Adding spatial diffusion in the 2-age model.}
\label{ssec:add_diffusion}

\begin{figure}[t!]
\centering
\subfigure[IBM]{\includegraphics[trim={0.1cm 0cm 0.5cm 0cm},clip, width=0.49\textwidth]{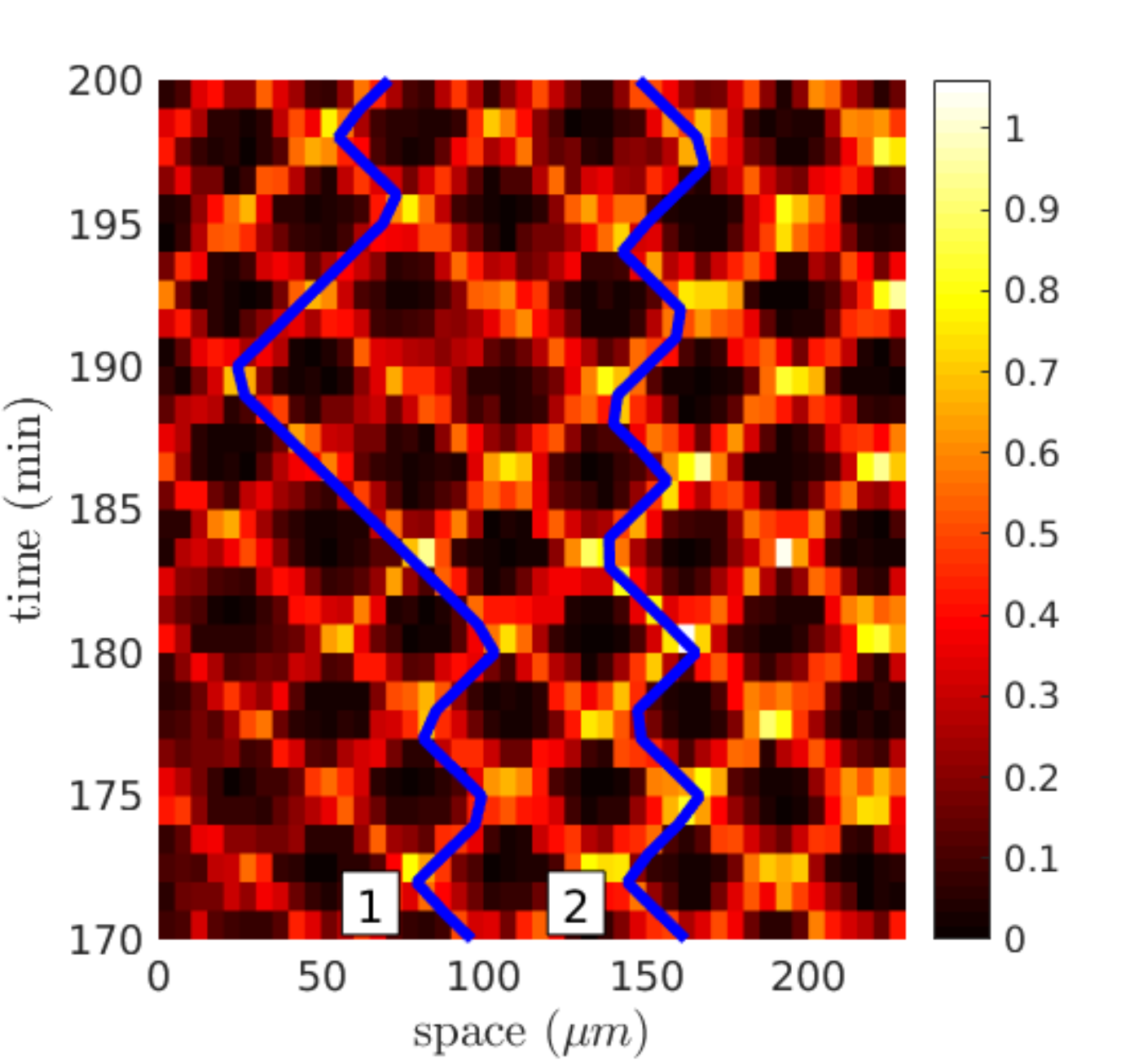}}
\subfigure[2-Age Model]{\includegraphics[trim={0.1cm 0cm 0.5cm 0cm},clip, width=0.49\textwidth]{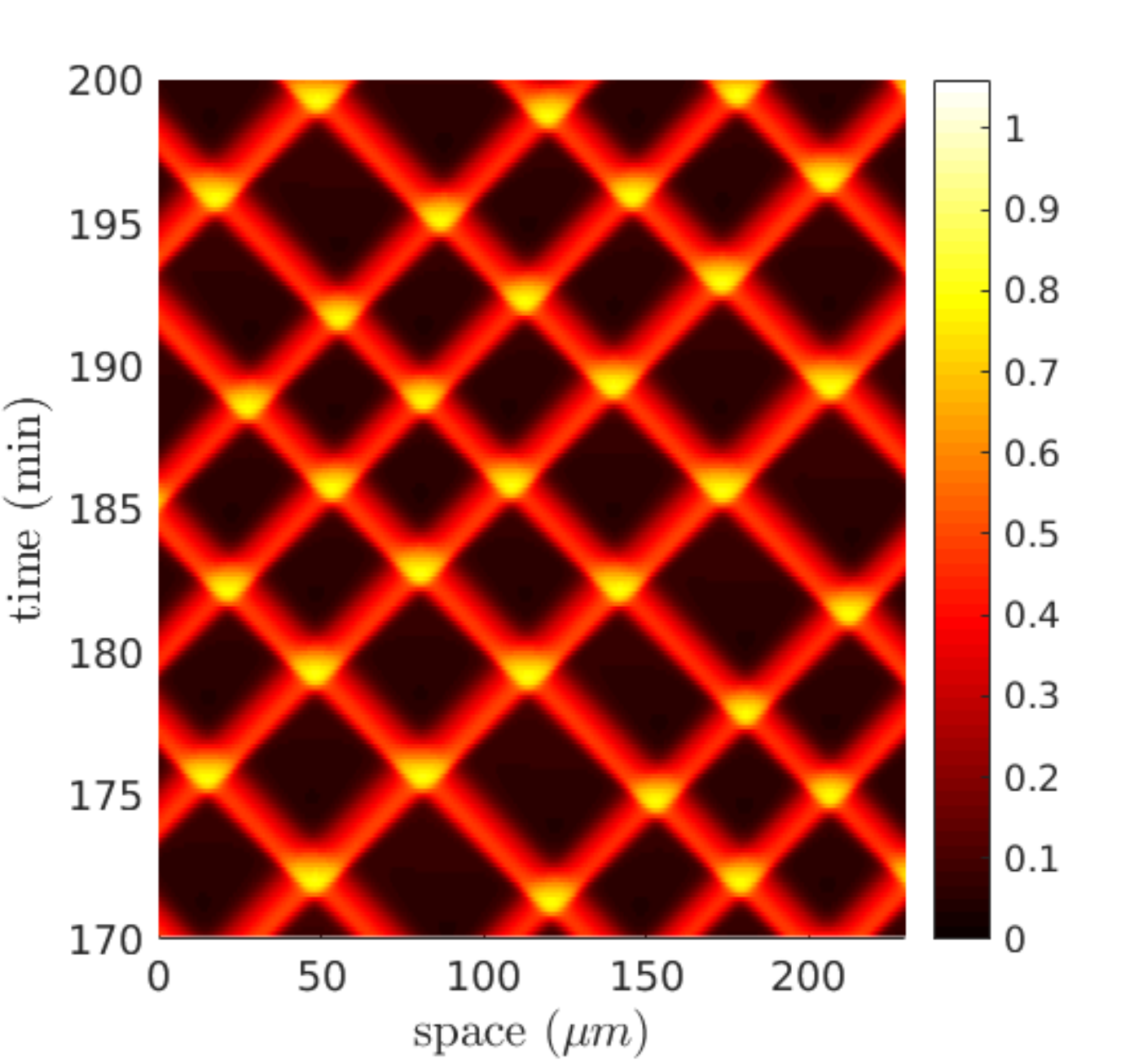}}
\caption{Comparison between IBM and 2-Age Model: Wave behavior. (a) shows the 1D densities along the rectangular strip marked in Fig.\,\ref{fig:IBM1}(a) calculated at each time step for 170 min $\leq t\leq 200$ min. It can be seen that waves persist over time, move with a speed of $\approx 9\,\mu m/$min and are not affected by collisions. Blue lines are the traces of the two particles marked in Fig\,\ref{fig:IBM1}(a). 
They mostly travel in wave crests, typically reverse upon wave collision and make almost no net movement along the strip.
(b) depicts the densities over the same time interval for the 1D 2-age model showing very similar macroscopic behavior. The units of both colorbars are bacteria per $\mu m^2$. }
\label{fig:IBM2}
\end{figure}

The above results show that the particle model provides a faithful approximation of the biological reality. Since particle models come with a high computational cost and are inherently hard to analyze analytically, we want to use the much simpler 2-age model to gain biological insight into wave formation. Motivated by the observation that alignment leads to a \emph{global} nematic mean direction, we assume $\bar\theta\equiv 0$ in the 2-age model, which amounts to setting $v(\bar\theta)=(1,0)^T$ and omitting the equation for $\bar\theta$. The remaining four equations constitute a non-linearly coupled system of transport-reaction equations with no spatial diffusion.\newline\par
\noindent
\paragraph{No wavelength control without spatial diffusion.} Simulating this system with initial conditions, randomly perturbed around the space-homogeneous steady state \eqref{eqn:steady}, we found that it produces counter-propagating density waves, just like its IBM counter-part. However, the wavelength can be arbitrarily small and is only limited by the diffusivity of the numerical scheme or the spatial resolution (see App.\,\ref{app:numerical_diffusion}). The waves in the IBM have a controlled wavelength, indicating that spatial diffusion is inherent in the IBM, but is ``lost'' during the derivation of the SOH. Rigorous derivation of the spatial diffusion term will be a topic for future work and will involve performing a Chapman-Enskog expansion \cite{chapman1970} as detailed e.g. in \cite{degond2010}. The fact that we need a higher order approximation, is most likely a consequence of the fact that for this system the microscopic scale (size of one bacterium, $\approx 5\,\mu m$) is small, but not negligibly small, compared to the size of the developing waves (order of 50-100$\mu m$). In this work we add a diffusion term of size $\delta$ to \eqref{eqn:macro_2age_diffFree} and use the IBM to estimate this otherwise difficult to measure parameter by fitting the produced wavelengths. In all of the following simulations we use the following diffusion-corrected version of \eqref{eqn:macro_2age_diffFree}:
\begin{subequations}
\label{eqn:macro_2age}
\begin{align}
& \pd_t \rho^0_+ + d_1 v_0\nabla_x\cdot (\rho^0_+ v(\bar \theta))=\delta \Delta\rho_+^0-\frac{1}{T}\rho^0_++\lambda(\sigma_+)\rho_-^1,\\
& \pd_t \rho^1_+ + d_1 v_0\nabla_x\cdot (\rho^1_+ v(\bar \theta))=\delta \Delta\rho_+^1+\frac{1}{T}\rho^0_+-\lambda(\sigma_-)\rho_+^1,\\
& \pd_t \rho^0_- - d_1 v_0 \nabla_x\cdot (\rho^0_- v(\bar \theta))=\delta \Delta\rho_-^0-\frac{1}{T}\rho^0_-+\lambda(\sigma_-)\rho_+^1,\\
& \pd_t \rho^1_- - d_1 v_0\nabla_x\cdot (\rho^1_- v(\bar \theta))=\delta \Delta\rho_-^1+\frac{1}{T}\rho^0_--\lambda(\sigma_+)\rho_-^1,
\end{align}
\end{subequations}

\noindent
\paragraph{The diffusion corrected 2-age model is a good approximation of the particle model.} 
We simulate  \eqref{eqn:macro_2age} with randomly perturbed, constant initial conditions of equal mean density as for Fig.\,\ref{fig:IBM1} (see caption of Fig.\,\ref{fig:IBM1} for details). We used $\delta=0.8\,\mu m^2/$min, fitted to produce the correct wavelength. In the 2-age model, opposing traveling waves also emerge (discussed in more detail in Sec.\,\ref{ssec:wildtype}). Fig.\,\ref{fig:IBM1}(b) compares the 1D densities calculated from the IBM to those of the 1D 2-age model: Crest and trough widths match closely. Further one can observe that the composition of the crests in terms of refractory and non-refractory bacteria match very well prior and after crest collisions (boxes A and B in the middle and lower rows). Finally also the macroscopic wave behavior is very similar to that observed in the space-time plots shown in Fig.\,\ref{fig:IBM2}. The remainder of the section is therefore devoted to analyzing the diffusion corrected 1D 2-age model \eqref{eqn:macro_2age}.

%%%%%%%%%%
% WAVE EMERGENCE
%%%%%%%%%%

\subsection{The wildtype: emergence of waves}
\label{ssec:wildtype}

\begin{figure}[t!]
\centering
\includegraphics[width=\textwidth]{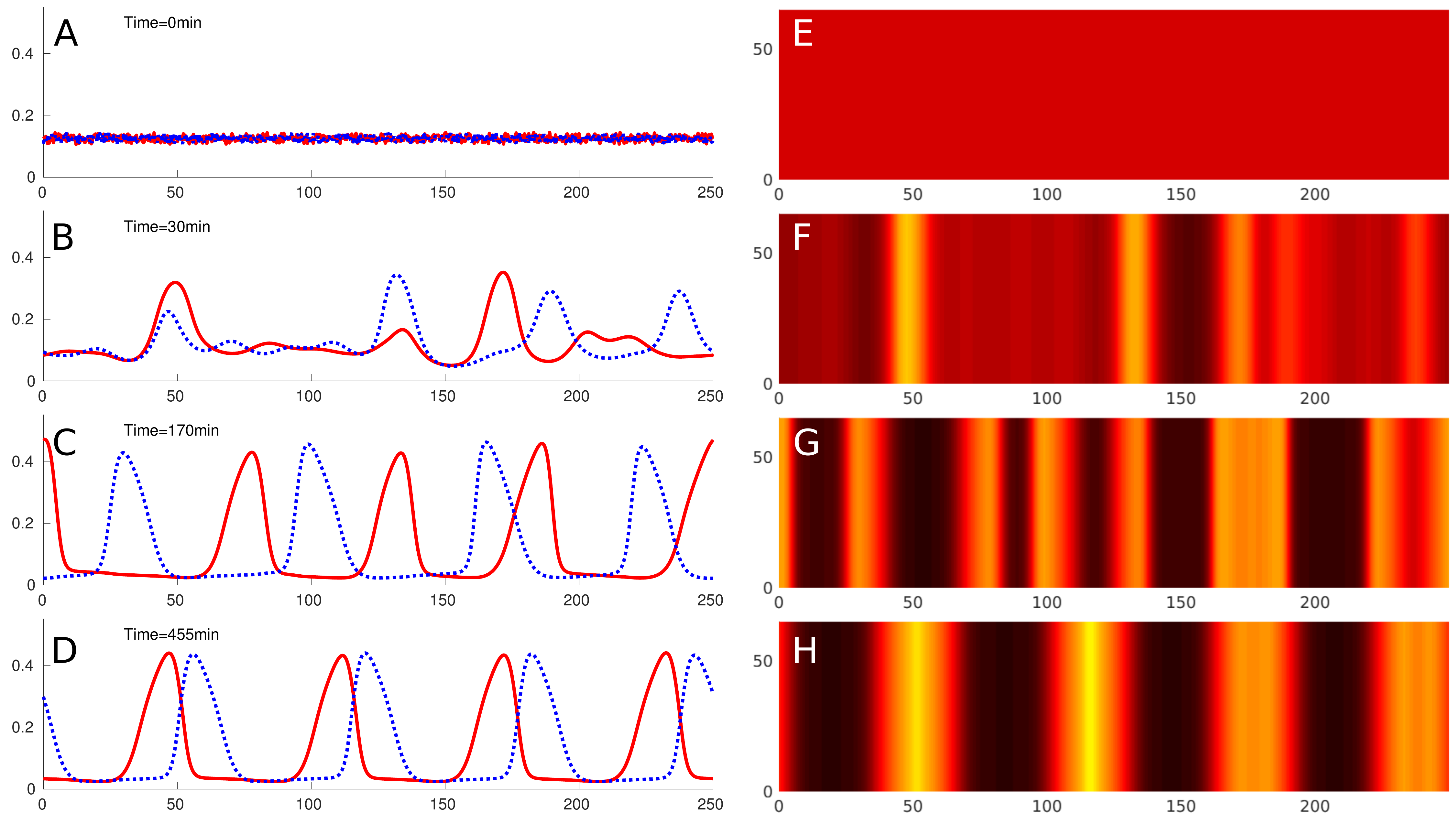}
\caption{Ripple Emergence. A-D: Time snapshots of the densities $\sigma_\pm$ of right-moving bacteria (red-solid) and left-moving bacteria (blue-dotted). The x-axis is in $\mu m$ and the y-axis is in bacteria per $\mu m^2$. E-H: "Microscopy-view", from above. The total local density of bacteria $\sigma_++\sigma_-$ is shown in 2D space at the same time points as A-D. All length units are $\mu m$.}
\label{fig:waveEmergence}
\end{figure}

We want to understand wave formation, shape and behavior in more detail using the 1D 2-age model \eqref{eqn:macro_2age}. All simulations are performed with periodic boundary conditions on an interval of length $L=250\,\mu m$ using the wildtype parameters listed in Table \ref{tab:parameters} (unless stated differently). The numerical method is discussed in App.\,\ref{app:numerics}. 
Note that system \eqref{eqn:macro_2age} conserves the total mass and we define $2\,m_0$ as the average total density, which is constant in time:
\begin{align*}
2\,m_0:=\frac{1}{L}\int_0^L \left(\rho_+^0+\rho_-^0+\rho_+^1+\rho_-^1\right)\dd x.
\end{align*}
As initial conditions we determine spatially uniform steady-state solutions by finding values which make the reaction terms on the right-hand side of \eqref{eqn:macro_2age} zero, yielding
\begin{align}
\label{eqn:steady}
\rho_+^0=\rho_-^0\equiv \frac{m_0\,\lambda(m_0)}{1/T+\lambda(m_0)}, \quad \rho_+^1=\rho_-^1\equiv \frac{m_0\,1/T}{1/T+\lambda(m_0)},
\end{align}
and perturb them with a uniform random distribution. These steady-state solutions reflect the fractions of non-refractory and refractory cells in the absence of spatial patterning. Large reversal rates, i.e. large values of $\lambda(m_0)$ will increase the fraction of refractory cells, because cells will spend less time on average in a non-refractory state. Small refractory periods $T$ on the other hand decrease them, as they will shortly become sensitive to C-signaling again.\newline\par
\begin{figure}[h!]
\centering
\includegraphics[width=\textwidth]{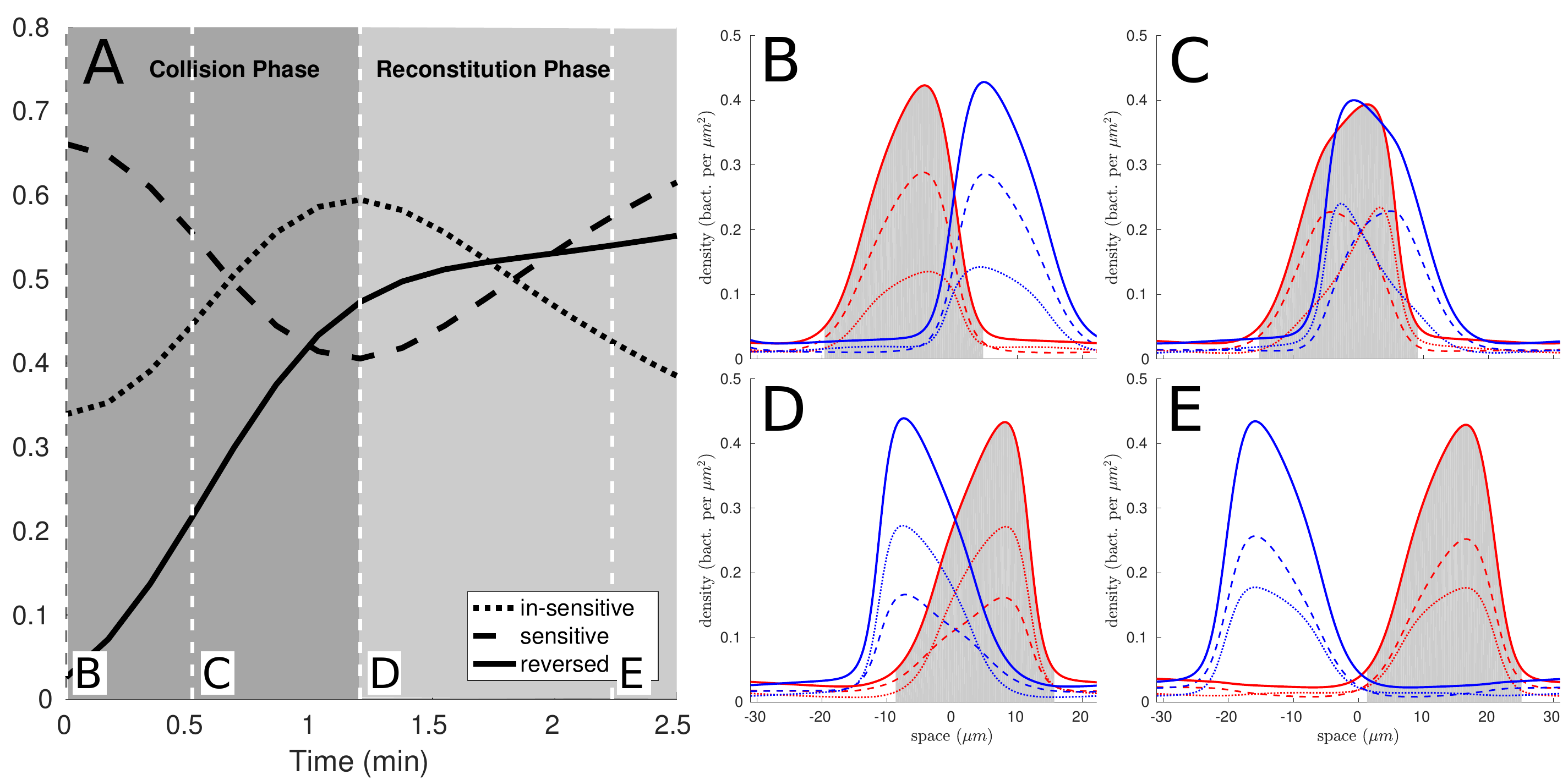}
\caption{Collision Study. A: The fraction of refractory (dotted) and non-refractory (dashed) bacteria in the wave over the time of one collision. The solid line shows the cumulative, total number of reversing bacteria in that wave as a fraction of the total number of bacteria originally present. The extent of the wave crest is defined by $\sigma_\pm\geq 0.05$ (see gray shaded regions in Fig. B-E on the right). White dashed lines and boxed letters mark the time points that are shown to the right. 
B-E: Time snapshots showing densities of refractory (dotted) and non-refractory (dashed) cells as well as their sum (solid) for the left- and right-moving groups (blue and red respectively). Parameters are listed in Tab.~\ref{tab:parameters}.}
\label{fig:collisionStudy}
\end{figure}

\noindent
In Fig.\,\ref{fig:waveEmergence} a time series of one simulation is shown. 
After about one hour bands of oppositely moving ripples start to develop and are fully established after two hours, after which their general shape and speed do not change anymore. The ripple crests move with a speed close to the bacterial speed ($9\,\mu m$/min) to the left and right respectively.  
The density ratio between crests and troughs is about $10$, which corresponds to the experimental values found e.g. in \cite{Sager1994}. Upon collisions of two such waves, the total bacteria densities (Fig.\,\ref{fig:waveEmergence}E-H) double, as described also in experiments. 
The shapes of the individual waves of left- and right-moving bacteria themselves seem to be almost unaffected by the collision, with only slight deformations. However, when inspecting the composition of the ripple crests in terms of refractory and non-refractory bacteria during a collision, one can observe two distinct phases: a \textit{collision phase} and a \textit{reconstitution phase} (see Fig.\,\ref{fig:collisionStudy}): while before the collision the fraction of refractory bacteria is low, it increases rapidly in the collision phase, indicating a high number of reversals taking place. 
In the reconstitution phase that follows, this fraction decreases again and resumes its original value. 
During this phase, the C-signal insensitive bacteria go through their refractory period and ``age" back into C-signaling sensitive bacteria. 
To estimate the fraction of bacteria that reverse during a collision, we compare the total number of reversing bacteria in one wave in the course of one collision to the original number of bacteria present in that wave before the collision. 
Fig.\,\ref{fig:collisionStudy} (left, solid line) shows that more than half of all bacteria reverse.

%%%%%%%%%%
% WHAT DOES T DO?
%%%%%%%%%%

\subsection{The influence of the refractory period $T$}
\paragraph{A memory-free model still produces traveling waves.}
A crucial part of the presented model is the introduction of the refractory period $T$. 
What is its influence on the bacterial behavior predicted by the 2-age model? In order to assess this, we investigate what changes in the absence of a refractory period, i.e. for $T\rightarrow 0$. The corresponding model can be interpreted as a \emph{memory-free} model, as bacteria retain no information about their previous reversals (see explanation below). Mathematically this can be realized by taking the limit $T\rightarrow 0$, i.e. bacteria are susceptible to C-signaling immediately after they reverse. 
System \eqref{eqn:macro_2age} then reduces to $\rho_\pm^0\equiv 0$ and $\rho_\pm^1=\sigma_\pm$ fulfilling (compare with Fig.\,\ref{fig:reaction_schematic}(c) for the reaction diagram)
\begin{subequations}
\label{eqn:macro_noClock}
\begin{align}
& \pd_t \rho^1_+ + d_1 v_0\nabla_x\cdot (\rho^1_+ v(\bar \theta))=\delta \Delta \rho^1_++\lambda(\rho^1_+)\rho^1_--\lambda(\rho^1_-)\rho^1_+,\\
& \pd_t \rho^1_- - d_1 v_0\nabla_x\cdot (\rho^1_- v(\bar \theta))=\delta \Delta \rho^1_-+\lambda(\rho^1_-)\rho^1_+-\lambda(\rho^1_+)\rho^1_-,
\end{align}
\end{subequations}
\begin{align}
\label{eqn:macro_noClock_theta}
&(\rho^1_++\rho^1_-)\pd_t \bar\theta+d_2v_0(\rho^1_+-\rho^1_-)(v(\bar\theta)\cdot \nabla_x)\bar\theta+\mu v_0\,v(\bar\theta)^\perp\cdot\nabla_x(\rho^1_+-\rho^1_-)=0.
\end{align}
\begin{remark}
\textit{The formally correct way to derive this system requires two steps: Firstly, system \eqref{eqn:macro_2age} needs to be non-dimensionalized by introducing a suitable reference timescale $\bar t$. A natural choice is $\bar t=\frac{L}{d_1 v_0}\approx 27$ min, i.e. the time it would take a non-reversing bacteria to cross the domain. The above limit then amounts to saying that $\e=T/\bar t$ is small compared to this typical time scale. Secondly, one needs to Taylor expand the functions $\rho_\pm^0$ and $\rho_\pm^1$ in terms of $\e$. 
One then finds that to the first order $\rho_\pm^0$ are zero and $\rho_\pm^1$ fulfill system \eqref{eqn:macro_noClock}.}
\end{remark}
\paragraph{}
The diffusion free version of this system has already been described in \cite{Degond2016}, where a memory-free myxobacteria model without an internal age variable $s$ was derived. 
Note that without the reaction term (i.e. $\lambda\equiv 0$), the system describes the macroscopic limit of purely nematic interactions, a phenomena of great interest in physics and studied in various works \cite{Degond_Navoret_2015,Ngo2012,Peruani2011}. 
Assuming a constant nematic direction $\bar\theta$,  the system reduces to two coupled transport-reaction equations; equations of this type were examined in the context of pattern formation and aggregation in biological systems e.g. in \cite{Eftimie2012,Lutscher2002} (see also the discussion below).\newline\par
We simulate the memory-free model \eqref{eqn:macro_noClock} in one space dimension (assuming $\bar\theta\equiv 0$), with the same parameters as for the 2-age model (Tab.\,\ref{tab:parameters}) and observe that the constant steady state again destabilizes under randomly perturbed initial conditions and traveling bumps occur. 
However, their widths vary greatly and do not seem to be controlled by the dynamics at all. In fact in \cite{Lutscher2002} it was commented that without diffusion the system seems to converge to piecewise constant traveling waves, traveling precisely at speed $d_1v_0$, which fulfill \eqref{eqn:macro_noClock} in a weak sense.
\begin{figure}[h!]
\centering
\includegraphics[width=0.85\textwidth]{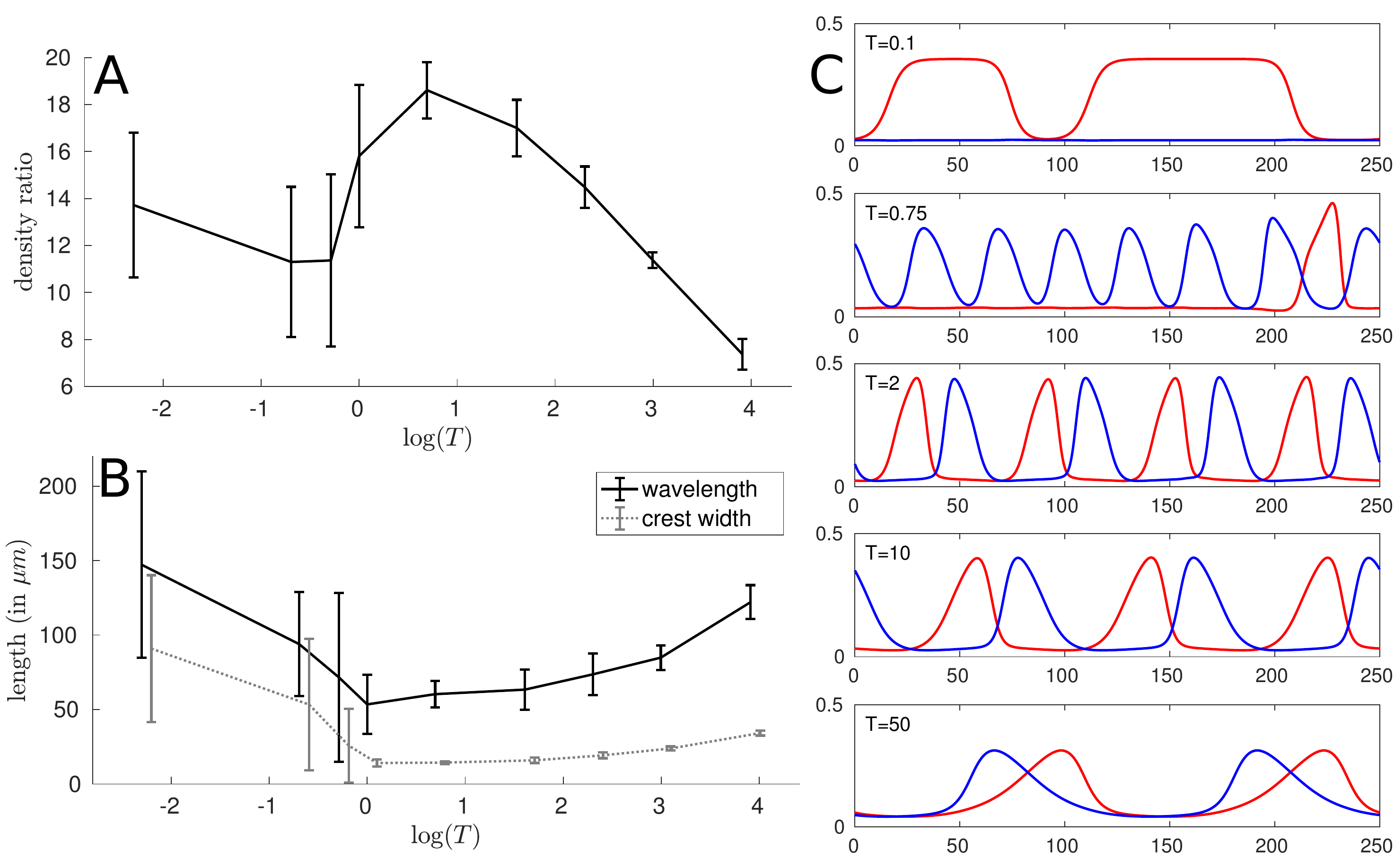}
\caption{Varying the refractory period $T$. Shown are the dependencies of the density ratio between wave crest and wave trough (A) and the wavelength and crest width (B) on the refractory period $T$. Error bars show $\pm$ standard deviation for a sample consisting of all waves from ten simulation runs per $T$. 
In the case of non-periodic waves, wavelengths were measured as distances between the center of masses of subsequent crests. 
C: Four examples of traveling wave profiles for different values of $T$. Shown are the total densities $\sigma_+$ (red-solid) and $\sigma_-$ (blue-dashed). Measured are the shapes after the system has reached equilibrium. Other parameters are listed in Tab.\,\ref{tab:parameters}.}
\label{fig:varyTr}
\end{figure}
\newline\par\noindent
\paragraph{The refractory period causes wave synchronization.} The results from the memory-free model suggest that while the refractory period might not be necessary for the formation of opposing traveling waves, it seems to be responsible for controlling the width of the individual traveling crests and synchronizes the waves by controlling the wavelength. 
To examine this further we systematically vary $T$ in the 2-age model and examine the resulting effects. In Fig.\,\ref{fig:varyTr} it can be observed that there are two parameter regimes. For very small refractory periods, the system behaves similarly to the memory-free, limiting system \eqref{eqn:macro_noClock}, i.e. while traveling waves occur, the crest width and wavelength varies greatly, indicating a lack of synchronization. After reaching a critical value of $T\approx 1$ min the waves become more synchronized and the crests share a common width. In this regime larger refractory periods lead to wider crests with larger wavelengths. How can we interpret these results?\newline\par
\begin{figure}[h!]
\centering
\includegraphics[width=\textwidth]{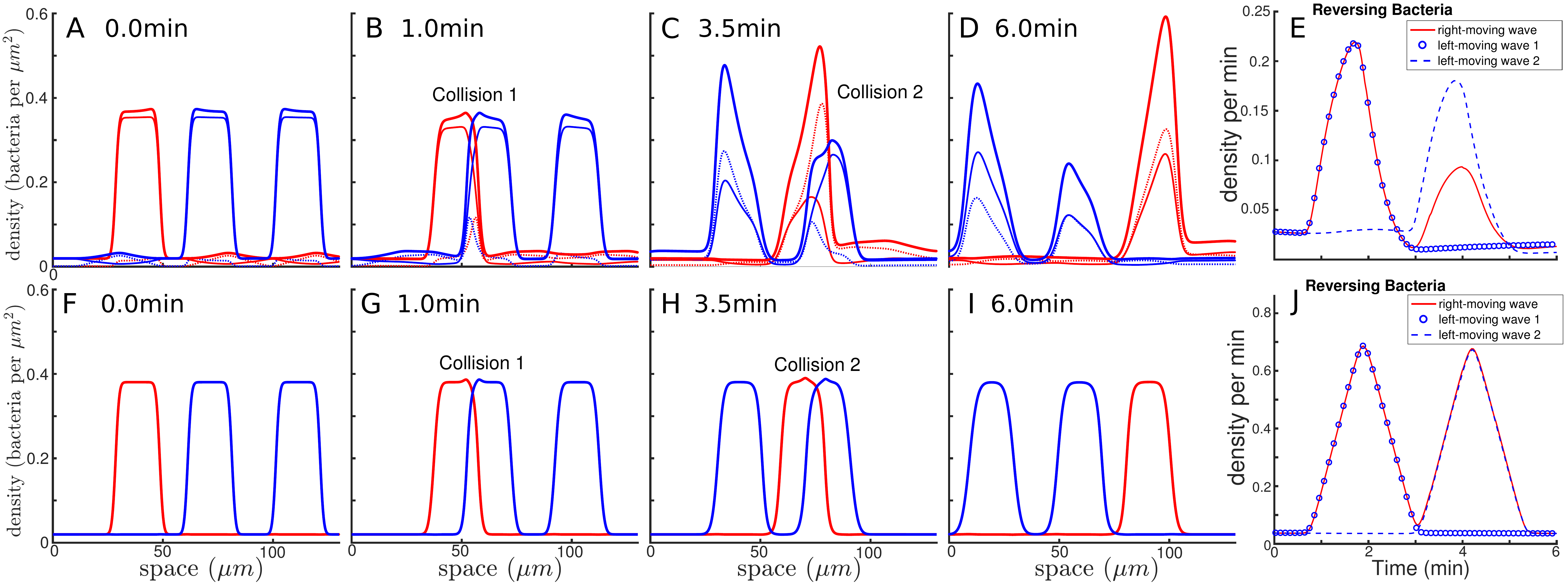}
\caption{Wavelength control. A-D and F-I: Four time snapshots from a 1D-simulation with refractory period (i.e. system \eqref{eqn:macro_2age}, A-D) and without (i.e. system \eqref{eqn:macro_noClock}, F-I). 
Shown are densities of non-refractory cells (thin-solid), refractory cells (thin-dashed), as well as their sums (thick-solid). Right- and left-moving densities are shown in red and blue, respectively. E and J: Average number of reversing bacteria per time in each of the three waves for the 2-age model (E) and the memory-free model (J) as a function of time, i.e. the average of $\lambda(\sigma_-)\rho_+^1$ (red-solid for right moving wave) and $\lambda(\sigma_+)\rho_-^1$ (blue-circles for leading left-moving wave and blue-dashed for following left-moving wave) over the whole wave. Parameters as in Tab.~\ref{tab:parameters} with $T=4$ min for the 2-age model.}
\label{fig:explainTr}
\end{figure}

\noindent
\paragraph{The refractory period imposes a time-delay between successive collisions.} The reason for the lack of synchronization in the case of small $T$ lies in the fact that bacteria retain no memory of previous reversals, hence two waves cannot affect each other. To understand how the refractory period enhances wave synchronization, it is instructive to look again at Fig.\,\ref{fig:collisionStudy}: as described above, during the reconstitution phase after a collision, the number of non-refractory cells grows back to its equilibrium value. 
This growth depends on $T$ (the smaller $T$, the faster the growth). 
If now the crest meets another crest before sufficiently many bacteria have regained their sensitivity to the C-factor, the wave will not be fully reflected off the oncoming wave. Fig.\,\ref{fig:explainTr} depicts the result of a simulation, in which one right-moving wave meets two left-moving waves traveling with a short wavelength for both the 2-age model (A-D) and the memory-free model (F-I). Fig.\,\ref{fig:explainTr}E and J depict how many bacteria reverse on average in each wave per minute (i.e. to be precise, if $x_a$ and $x_b$ mark the beginning and end positions of a wave, Fig.\,\ref{fig:explainTr}E/J depict $\frac{1}{x_b-x_a}\int_{x_a}^{x_b} \lambda(\sigma_\mp)\rho_\pm^1 \dd x$ in both models). At the first collision the same (high) number of bacteria reverse in both waves, the waves are reflected off each other and the number of reversing bacteria is similar for both the 2-age model and the memory-free model. In the second collision for the 2-age model, however, the right-moving wave contains much fewer non-refractory cells (i.e. $\rho_+^1$ is lower) than the second left-moving wave, hence it is only partially reflected. The overall effect is a significant reduction of the second left-moving wave. For the memory-free case, the second collision resembles exactly the first collision, hence the second wave is unchanged.

%%%%%%%%%%
% WAVE FORMATION
%%%%%%%%%%

\subsection{Parameter-dependence of wave formation}

\paragraph{Two necessary wave conditions.} In many experimental set-ups, one of the main output is whether or not myxobacteria colonies form waves. We therefore examine what parameter combinations lead to wave formation.  %By non-dimensionalizing system \eqref{eqn:macro_2age} using as characteristic lengthscale the domain size $\bar x=L$, as characteristic timescale $\bar t=\frac{d_1v_0}{L}$, the time it would take a non-reversing bacteria to cross the domain and as characteristic density the mean total density $\bar\sigma=m_0$, it can easily be seen that the system is characterized by only four dimensionless quantities
%\begin{align*}
%\tau=\frac{T}{\bar t}, \quad \underline\lambda=\lambda_m\bar t,\quad \overline \lambda=\lambda_M\bar t,\quad r=\frac{\bar\rho}{m_0}
%\end{align*}
The previous section suggests that the length of the refractory period $T$ has no influence on the appearance of waves. We therefore perform a rough parameter scan over the shape parameters of the reversal function $\lambda(\rho)$, i.e. the spontaneous reversal rate $\lambda_m$, the maximal reversal rate $\lambda_M$ and the inflection density $\bar\rho$. Rigorous mathematical stability analysis of system \eqref{eqn:macro_2age} will give more insight into the precise stability regions also for other shapes of $\lambda(\rho)$ and will be the subject of future work. However, several conclusions can already be drawn. We find two wave formation conditions:\newline
\textbf{Condition A:} The maximal reversal rate needs to be large enough compared to the spontaneous reversal rate ($\lambda_M>3\lambda_m$).
\newline \textbf{Condition B:} The inflection density and the average total density need to be of similar order ($\bar\rho\approx m_0$). \newline How can we links this to experiments?\newline
\begin{remark}
\textit{A non-dimensionalization of the system shows that its behavior depends on four dimensionless quantities
\[
\frac{d_1v_0 T}{L}, \quad \frac{\lambda_m L}{d_1 v_0}, \quad \frac{\lambda_M L}{d_1 v_0}\,\,\, \text{and}\,\,\, \frac{\bar\rho}{m_0}.
\]
The first quantity compares the time it takes a just reversed bacterium to age back into a reversible state, to the time it takes a (non-reversing) bacterium to cross the domain. The next two quantities can be interpreted as the expected number of reversals during one domain crossing for reversal rates $\lambda_m$ and $\lambda_M$ in the absence of a refractory period. The last quantity compares the initial average density to the inflection density, i.e. the density of opposing bacteria, where the reversal function $\lambda$ is the most sensitive to changes in density. Examining $\lambda_m$, $\lambda_M$ and $\bar\rho$ therefore amounts to analyzing the last three quantities.}
\end{remark}
\noindent
\paragraph{Mutation experiments: hypo- and hyper-reversing bacteria.}
To assess how the reversal behavior impacts the ripple formation, Sager and Kaiser have used M. xanthus strains, that have an insertion mutation in the \textit{frzCD} gene, which has been shown to impact the reversal probability \cite{Sager1994}. 
Isolated individuals of the \textit{hypo-reverses} change direction on average $\approx$10 times less frequent and \textit{hyper-reversers}  $\approx$5 times more often as compared to the wildtype. Both strains have lost the ability to form ripples and it is suggested that the mutation affects the spontaneous reversal rate. Wave condition (A) suggests that if the spontaneous reversal rate $\lambda_m$ is too large compared to the maximal reversal rate, no waves will form, in agreement with the hyper-reverser experiments. Our wave formation conditions do not explain the absence of waves in hypo-reversers, however measuring the reversal rate $\lambda(\rho)$ experimentally could shed light on this question.\newline\par
\begin{figure}[h!]
\centering
\includegraphics[width=0.8\textwidth]{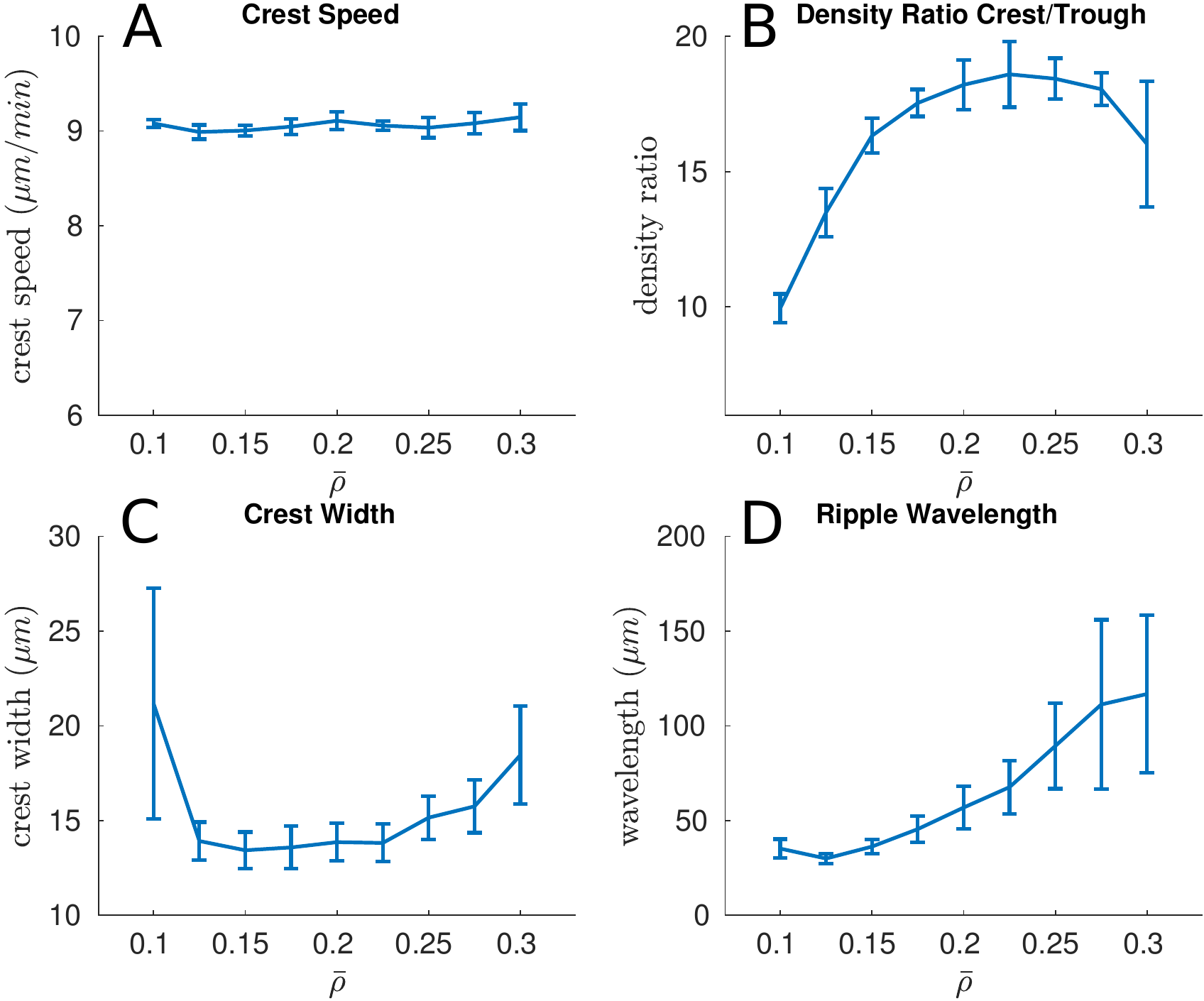}
\caption{Varying the inflection density $\bar\rho$. The effect of varying $\bar\rho$ on various wave characteristics. 
Depicted are mean $\pm$ standard deviation for all waves for 10 simulation runs per $\bar\rho$. 
Crest speeds were measured over the time that it takes for one wave to cross the simulation domain. Other parameters are listed in Tab.\,\ref{tab:parameters}.}
\label{fig:vary_rhoBar}
\end{figure}

\noindent
\paragraph{Dilution experiments: changing the fraction of C-signal competent cells.}
Another way to demonstrate the influence of C-signaling on the ripple formation was a second set of experiments performed by Kaiser and Sager \cite{Sager1994}: two strains of myxobacteria were used, a C-signaling competent wildtype strain $csgA^+$ and a mutant strain $csgA^-$, which can not produce C-signal, but can respond to it. By themselves ensembles of  $csgA^-$ cells do not have the ability to form ripples. In their experiments Kaiser and Sager changed the fraction of $csgA^+$ and $csgA^-$ cells and measured the changes in ripple wavelength, speed and width. 
To simulate this situation using our 2-age model, we note that, since $csgA^-$ cells react to C-signals in the same manner as the wildtype, the fraction of $csgA^+$ and $csgA^-$ will be constant everywhere. This was confirmed in \cite{Sager1994}. Let the fraction of C-signaling competent $csgA^+$ cells be $q\in[0,1]$. At some density of opposing bacteria $\sigma$, therefore only $q\sigma$ cells contribute to the amount of signal produced, i.e. $\lambda(\sigma_\pm)$ in system \eqref{eqn:macro_2age} is replaced by $\lambda(q\sigma_\pm)$. Substituting this into \eqref{eqn:lambda}, the expression for $\lambda$, we see that this in fact simply changes $\bar\rho$ to $\bar\rho/q$. In view of wave condition (B), this suggests that if $q$ is too small, this would inhibit wave formation. This is in agreement with the experiments described in \cite{Sager1994}, although it should be noted that for very small fractions of $csgA^+$ cells, for which waves were still present in the experiment, the corresponding large values of $\bar\rho$ do not produce waves anymore in our model.  
We further study how varying the inflection density $\bar\rho$ affects the wave shape (within the range that supports ripple formation). 
Fig.\,\ref{fig:vary_rhoBar} depicts the impact of this variation on various wave characteristics. Wave speed is not affected and stays close to the individual bacterial speed (Fig.\,\ref{fig:vary_rhoBar}A). For regular waves, crest width increases, as does ripple wavelength (Fig.\,\ref{fig:vary_rhoBar}C/D), both of which are in agreement with the findings in \cite{Sager1994}. 
Our model also predicts a non-monotonous dependence of the ratio of the number of cells in the crest to the number of cells in the trough as a function of $\bar \rho$ (Fig.\,\ref{fig:vary_rhoBar}B), which could easily be addressed experimentally.

%%%%%%%%%%%%%%%%%%%%%%%%%
%%%%%%%%%%%%%%%%%%%%%%%%%
%%%%
%%%%  	DISCUSSION
%%%%
%%%%%%%%%%%%%%%%%%%%%%%%%
%%%%%%%%%%%%%%%%%%%%%%%%%

\section{Discussion}
\label{sec:discussion}

We presented a novel, continuous, age-structured macroscopic model of myxobacteria, derived systematically from an individual-based model. The main assumptions of the model are nematic alignment, a density-dependent reversal function and a refractory period of fixed length, which introduces a memory effect.\newline\par

In excellent agreement with experimental data on myxobacteria, simulations of the full IBM show the development of periodic waves, traveling in opposing directions and being reflected upon collision. We performed an in-depth numerical investigation of the one-dimensional macro-model for the case of two sensitivity/age groups: refractory cells, incapable of reversing, and non-refractory cells, that can reverse and are sensitive to C-signaling. 
A main result of our analysis is that the refractory period is not responsible for wave formation, but for wave synchronization. This is because it introduces a memory effect, that controls wavelengths. The existence of a refractory period is known for example for \emph{D.~discoideum} \cite{shaffer1975}. 
The idea of a refractory period for myxobacteria has originally been brought forward in the model presented in \cite{Igoshin2001}, where it is suggested that myxobacteria also go through an insensitivity period following a reversal. 
Assuming that the length of the refractory period is unaffected by the density of opposing bacteria, we show that the memory effect introduced by a fixed refractory period is sufficient to explain ripple synchronization. 
We discovered two wave formation conditions that are consistent with experimental results: the maximal reversal rate needs to be large enough compared to the spontaneous reversal rate and the average density of the myxobacteria colony needs to be close to $2\bar\rho$ with $\bar\rho$ being the inflection density at which the reversal function reacts the most sensitively to density changes. This predicts that both very high and very low densities will inhibit wave formation in myxobacterial colonies.\newline
\par
A strength of the Self-Organized Hydrodynamics (SOH) models is that they are directly derived from the corresponding IBM by the method of Generalized Collision Invariants (GCI). 
This allows for easy and transparent inclusion of assumptions, such as the density-dependent reversal rate and the refractory period. The simulations presented in Sec.\,\ref{ssec:add_diffusion} show good agreement between the IBM and diffusion corrected 2-age model, however, to fully validate the correspondence between the macroscopic model and the IBM, a formal derivation of the mass diffusion term \cite{chapman1970,degond2010} will be necessary. The lack of control over the number of waves in absence of diffusion shows that for phenomena where the ratio between microscopic and macroscopic scales is not very small, higher order terms can be crucial and hence the additional effort for deriving them is justified.\newline
\par

In this work we concentrated on wave formation, where both experimental results and particle simulations suggest that the dynamics can be studied in one space dimension. However, several other macroscopic patterns are known for myxobacteria, most notably the formation of large aggregates. The combination of simulation results of both the IBM and the two-dimensional 2-age model as well as analytical results on the latter might shed light on what parameters cause cells to switch between a uniform state to ripple formation and aggregation. Several aspects will need to be considered: could the reversal frequency depend on both the densities of the opposing group as well as on the aligned one? In \cite{Lutscher2002} such cases were analyzed for a memory-free model and simulations also showed wave formation. Further the authors in \cite{Lutscher2002} also noted that small changes in the reversal function $\lambda(\rho)$ can cause the system to switch from ripples to aggregation. It will be interesting to extend the results to the age-structured model. In \cite{Jelsbak2003} experiments were described in which C-signaling and thereby local densities affect the gliding speed of bacteria. 
Such density-dependent parameters can easily be incorporated into SOH models and significantly impact the dynamics \cite{Frouvelle2012}.\newline\par

A large number of IBMs exist for self-propelled particles such as myxobacteria \cite{Sliusarenko2006} and detailed numerical and statistical analysis of their properties have significantly contributed to the understanding of emergent phenomena and phase transitions. While they allow for direct comparison between the experimental data about the behavior of the individuals, they are limited in terms of insights into macroscopic phenomena. 
For macro-systems such as the 2-age model on the other hand, stability or asymptotic  analysis can be performed which can elucidate precise parameter dependencies and long-term behavior. The macro-system of \cite{Igoshin2001} triggered a number of works examining wavelength determination \cite{Bonilla2016,Igoshin2004b}, demonstrating the potential insight gained through analytical methods. For the macro-system presented here, a rigorous analytical treatment will be the subject of future work. Several works deal with the linkage between IBMs and meso- or macro-models, e.g. both in \cite{Borner2002} and \cite{Grossmann2016} where continuous models are derived from particle models. The methods presented there as well as the GCI method offer a powerful option of combining the strengths of both particle and continuous based methods. 
\newline\par

Our work together with the results in \cite{Borner2002,Igoshin2001,Sliusarenko2006} strongly suggests the existence of a refractory period. 
Several aspects of our findings can be addressed experimentally: our main results is that wave formation and wave synchronization are independent phenomena, which would suggest separate molecular mechanisms. 
We therefore predict that mutants that form non-synchronized traveling waves of various width have a density-dependent reversal frequency, but no refractory period. 
As to the length of the refractory period, in \cite{Igoshin2001} experimental data argued for refractory periods of under 1 min. 
Our model argues that $T\approx 1$ min presents the lower limit for the formation of synchronized waves. Larger refractory periods also lead to periodic waves, but it takes the system much longer to produce steadily moving wave. Hence the value $T\approx 1$ min is the fastest way to make synchronized waves, presenting a possible, evolutionary answer as to why this particular refractory period length evolved.

\appendix

\section{Appendix}

%%%%%%%%%
%  NUMERICAL METHOD
%%%%%%%%%

\subsection{Numerical Methods}
 \label{app:numerics}
\paragraph{Particle Model.} The IBM of Sec.\,\ref{ssec:particle} was simulated in a square domain with periodic boundary conditions, using the method described in \cite{Motsch_Navoret_MMS11}.\newline\par\noindent 
\paragraph{2-Age Model.} We simulated the 2-age model \eqref{eqn:macro_2age} in one space dimension with periodic boundary conditions and $\bar\theta\equiv 0$. The transport and reaction terms were implemented using operator splitting with explicit upwind or downwind (for the $+$ and $-$ family respectively) finite differences for the transport term, an implicit finite difference scheme for the diffusion term and an explicit treatment of the reaction term. We also tested implementing the reaction term with an explicit Runge-Kutta (4,5) formula using the ode45 solver of Matlab and a Lax-Friedrichs Scheme for the transport term, both with no significant gains in accuracy. Tab.\,\ref{tab:parameters_numerics} shows the parameters used for both simulation set-ups.

%%%%%%%%%
%  WHY ADD DIFFUSION
%%%%%%%%%

\subsection{Controlling the number of waves.}

\begin{figure}[t!]
\centering
\includegraphics[width=\textwidth]{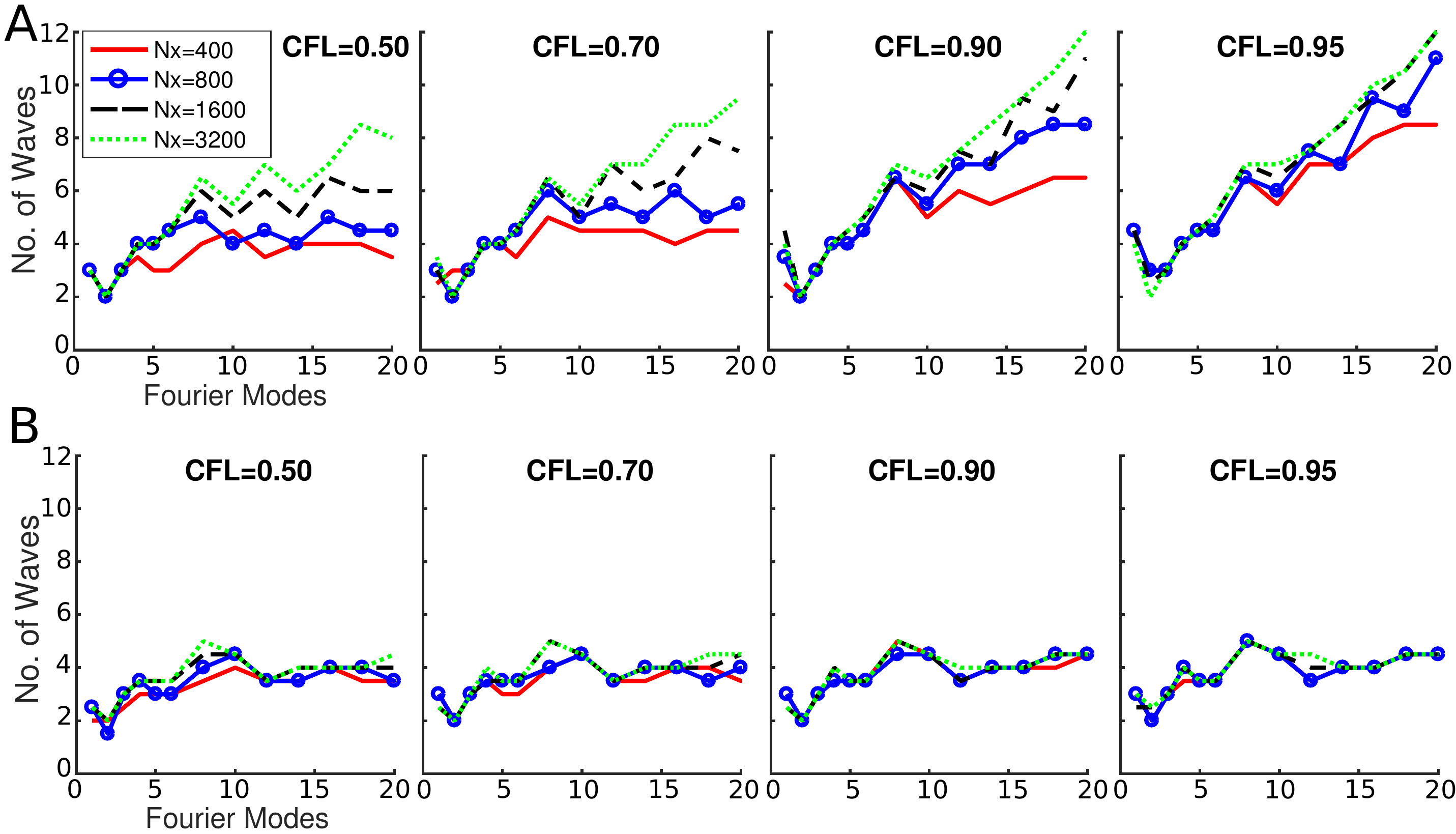}
\caption{Wave number determination. A: No diffusion. Depicted is the average number of peaks of $\sigma_-$ and $\sigma_+$ at time $ = 300$ min using the diffusion-free model \eqref{eqn:macro_2age_diffFree}, in dependence on the number of Fourier modes $K$. Varied were the CFL condition and the number of spatial discretization points Nx. B: With diffusion. As A, but using the diffusion-corrected model \eqref{eqn:macro_2age} with $\delta=0.8$. All other parameters were as in Tab.\,\ref{tab:parameters} and \ref{tab:parameters_numerics}.}
\label{fig:numerical_issues}
\end{figure}

\label{app:numerical_diffusion}
We want to asses whether \eqref{eqn:macro_2age_diffFree} encodes information about the number of waves \textit{naturally} created. We use as initial conditions small perturbations from \eqref{eqn:steady}, however using completely random data can be problematic due to its non-smootheness and because the result could be influenced by the spatial discretization. To avoid these problems, but mimic randomness, we use sums of Fourier-modes as perturbations:
\begin{align}
\sum_{k=1}^K a_k \cos{(2\pi k x)} +  b_k \sin{(2\pi k x)},
\end{align}
where $a_k, b_k$ are chosen randomly in $[-1,1]$. Finally we normalize the resulting perturbations so that the maximal amplitude doesn't exceed $\e=0.05$. For the time discretization we used $\Delta t=\Delta x/(d_1 v_0)\times $CFL, where CFL is the Courant-Friedrichs-Lewy number. The up-/down winding scheme for the transport terms implies that we need CFL$\leq1$ for stability. Without the reaction terms, the transport equations could be solved numerically exactly, if CFL=1. Then the numerical diffusion introduced by the discretization of the transport term increases for smaller CFL.\newline\par
\noindent
\paragraph{Arbitrary small wavelengths in the absence of diffusion.} To examine the effects of numerical diffusion and spatial resolution on the wavelength, we vary 1.) the maximal number of Fourier modes $K$ used in the initial conditions, 2.) the spatial discretization $\Delta x$ and 3.) the CFL condition. For each choice of $K$ we use the same initial conditions, i.e. the same random numbers $a_k, b_k$. Each simulation is run until time = $300$ min and the number of waves is determined as the average number of peaks of $\sigma_+$ and $\sigma_-$. Fig.\,\ref{fig:numerical_issues}A shows the outcome. For large $K$, for each fixed CFL number, a finer grid leads to more waves. On the other hand for a fixed grid size, the closer CFL is to $1$, the more waves we observe. This indicates that the upper bound of the number of waves is caused by the numerical diffusion and the spatial resolution, rather than by the equations themselves. However, the minimal number of waves seems to be encoded in the equations, as it is largely independent of the spatial discretization and CFL number. The issue of wavelength selection for a similar system was also discussed in \cite{scheel2017}.\newline\par

To test the influence of diffusion wavelengths, we repeat the same simulations, using the diffusion-corrected model \eqref{eqn:macro_2age}. Fig.\,\ref{fig:numerical_issues}B shows that now the behavior is consistent for different $\Delta x$ and the maximal number of waves is controlled. Since in the IBM we observe regular waves of relatively fixed wavelengths, we conclude that a small amount of density diffusion is inherent in the IBM and leads to a control in wavelength. This diffusion is lost during the derivation of the SOH model, but is necessary to avoid very small wavelengths (see discussion in Sec.\,\ref{ssec:add_diffusion}).

%%%%%%%%%
%  PARAMETERS
%%%%%%%%%

\begin{table}[t]
\begin{center}
\begin{tabular}{| c | c | c| c | }
  \hline			
  Name & Meaning & Value & Comment\\ 
  \hline
  $v_0$ & bacterial speed & $9\, \mu m /$min & \cite{Sager1994}\\
  $\lambda_m$ & spontaneous reversal rate & $0.07 /$min& \cite{Sager1994,Sliusarenko2006,Welch2001} \\
$\lambda_M$ & maximal reversal rate & $1.5 /$min& \cite{Sager1994,Sliusarenko2006,Welch2001} \\
$\bar\rho$ & inflection density & $0.2/\mu m^2$& fitting parameter \\
$T$ & refractory period & 2 min & estimates in \cite{Igoshin2001,Borner2002}\\
$D$ & angular diffusion constant & $0.1$/min& \multirow{2}{*}{ \Large{$\}$} \normalsize leads to $d_1=0.99$}\\
$\nu$ & alignment frequency & $100/$min& \\
$2\,m_0$ & total average density & $0.25 /\mu m^2$& \cite{Sliusarenko2006} \\
  \hline  
\end{tabular}
 \caption{Biological Parameters}
 \label{tab:parameters}
\end{center}
\end{table}

\begin{table}[t]
\begin{center}
\begin{tabular}{| c | c | c| }
  \hline			
  Name & Meaning & Value \\ 
  \hline
\multicolumn{3}{|c|}{Simulation parameters of IBM}\\
\hline
  $L_x,L_y$ & width and length of simulation domain & $L_x=L_y=250\,\mu m$   \\
  $N$ & total number of bacteria & $2\,m_0\,L_x\,L_y=15625$\\
$\Delta t$ & time step & $0.01$ min  \\
$R$ & interaction radius & $5\,\mu m$\\
   \hline
\multicolumn{3}{|c|}{Simulation parameters of 1D 2-age model}\\
\hline
  $L$ & simulation domain & $250\,\mu m$   \\
$\Delta x$ & spatial step & $0.313\,\mu m$  \\
$\Delta t$ & time step for transport operator & $\Delta x/(d_1v_0)0.95=0.033$ min\\
$\delta$ & diffusion constant & $0.8\,\mu m^2/$ min\\
   \hline
\end{tabular}
 \caption{Numerical Parameters}
 \label{tab:parameters_numerics}
\end{center}
\end{table}

\section*{Acknowledgements}
PD acknowledges support by the Engineering and Physical Sciences 
Research Council (EPSRC) under grant no. EP/M006883/1, by the Royal 
Society and the Wolfson Foundation through a Royal Society Wolfson 
Research Merit Award no. WM130048 and by the National Science 
Foundation (NSF) under grant no. RNMS11-07444 (KI-Net). PD is on leave 
from CNRS, Institut de Math\'ematiques de Toulouse, France. AM was supported by the Austrian Science Fund (FWF) through the doctoral school \emph{Dissipation and Dispersion in Nonlinear PDEs} (project W1245) as well as the Vienna Science and Technology Fund (WWTF) (project LS13/029). HY acknowledges the support by Division of Mathematical Sciences [KI-Net NSF RNMS grant number 1107444]; DFG Cluster of Excellence \emph{Production technologies for high-wage countries} [grant number DFG STE2063/1-1], [grant number HE5386/13,14,15-1]. HY and AM gratefully acknowledge the hospitality of the Department of Mathematics, Imperial College London, where part of this research was conducted. 

\section*{Data availability}
No new data were collected in the course of this research.

\section*{Supplmentary information}
\href{https://figshare.com/articles/IBM_video_avi/5769408}{Link to video on figshare}

\bibliographystyle{siam}
\bibliography{myxoclock_bibliography}

\begin{thebibliography}{10}

\bibitem{Baskaran2008}
{\sc A.~Baskaran and M.~Marchetti}, {\em Hydrodynamics of self-propelled hard
  rods}, Phys. Rev. E, 77 (2008), p.~011920.

\bibitem{bellomo2012}
{\sc N.~Bellomo and J.~Soler}, {\em On the mathematical theory of the dynamics
  of swarms viewed as complex systems}, Math. Models Methods Appl. Sci., 22
  (2012), p.~1140006.

\bibitem{Bertin2009}
{\sc E.~Bertin, M.~Droz, and G.~Gr\'{e}goire}, {\em Hydrodynamic equations for
  self-propelled particles: microscopic derivation and stability analysis}, J.
  Phys. A: Math. Theor., 42 (2009), p.~445001.

\bibitem{blanchet2016topological}
{\sc A.~Blanchet and P.~Degond}, {\em Topological interactions in a
  {B}oltzmann-type framework}, J. Stat. Phys., 163 (2016), pp.~41--60.

\bibitem{bolley2012mean}
{\sc F.~Bolley, J.~A. Ca{\~n}izo, and J.~A. Carrillo}, {\em Mean-field limit
  for the stochastic {V}icsek model}, Appl. Math. Lett., 25 (2012),
  pp.~339--343.

\bibitem{Bonilla2016}
{\sc L.~L. Bonilla, A.~Glavan, and A.~Marquina}, {\em Wavelength selection of
  rippling patterns in myxobacteria}, Phys. Rev. E, 93 (2016).

\bibitem{Borner2002}
{\sc U.~B\"{o}rner, A.~Deutsch, H.~Reichenbach, and M.~B\"{a}r}, {\em Rippling
  patterns in aggregates of myxobacteria arise from cell-cell collisions},
  Phys. Rev. Lett., 89 (2002), p.~078101.

\bibitem{carlen2015boltzmann}
{\sc E.~Carlen, M.~C. Carvalho, P.~Degond, and B.~Wennberg}, {\em A {B}oltzmann
  model for rod alignment and schooling fish}, Nonlinearity, 28 (2015),
  p.~1783.

\bibitem{Carrillo2009}
{\sc J.~Carrillo, M.~D\textsc{\char13}Orsogna, and V.~Panferov}, {\em Double
  milling in self-propelled swarms from kinetic theory}, Kinet. Relat. Models,
  2 (2009), pp.~363--378.

\bibitem{chapman1970}
{\sc S.~Chapman and T.~G. Cowling}, {\em The mathematical theory of non-uniform
  gases: an account of the kinetic theory of viscosity, thermal conduction and
  diffusion in gases}, Cambridge university press, 1970.

\bibitem{Chate2008}
{\sc H.~Chat\'{e}, F.~Ginelli, G.~Gr\'{e}goire, and F.~Raynaud}, {\em
  Collective motion of self-propelled particles interacting without cohesion},
  Phys. Rev. E, 77 (2008), p.~046113.

\bibitem{Chuang2007}
{\sc Y.~Chuang, M.~D\textsc{\char13}Orsogna, D.~Marthaler, A.~Bertozzi, and
  L.~Chayes}, {\em State transitions and the continuum limit for a 2{D}
  interacting, self-propelled particle system}, Phys. D, 232 (2007),
  pp.~33--47.

\bibitem{Degond_Delebecque_Peurichard_2015}
{\sc P.~Degond, F.~Delebecque, and D.~Peurichard}, {\em Continuum model for
  linked fibers with alignment interactions}, Math. Models Methods Appl. Sci.,
  26 (2016), pp.~269--318.

\bibitem{degond2016new}
{\sc P.~Degond, A.~Frouvelle, and S.~Merino-Aceituno}, {\em A new flocking
  model through body attitude coordination}, Math. Models Methods Appl. Sci.,
  27 (2017), pp.~1005--1049.

\bibitem{Degond_Liu_2012}
{\sc P.~Degond and J.-G. Liu}, {\em Hydrodynamics of self-alignment
  interactions with precession and derivation of the
  {L}andau-{L}ifschitz-{G}ilbert equation}, Math. Models Methods Appl. Sci., 22
  Suppl. 1 (2012), p.~114001.

\bibitem{Degond_Liu_Ringhofer_2014}
{\sc P.~Degond, J.-G. Liu, and C.~Ringhofer}, {\em Evolution of wealth in a
  nonconservative economy driven by local {N}ash equilibria}, Philos. Trans. A,
  372 (2015), p.~20130394.

\bibitem{Degond2016}
{\sc P.~Degond, A.~Manhart, and H.~Yu}, {\em {A continuum model for nematic
  alignment of self-propelled particles}}, Discrete Contin. Dyn. Syst. Ser. B,
  22 (2017), pp.~1295--1327.

\bibitem{Degond2008}
{\sc P.~Degond and S.~Motsch}, {\em Continuum limit of self-driven particles
  with orientation interaction}, Math. Models Methods Appl. Sci., {\bf 18}
  (2008), pp.~1193--1215.

\bibitem{Degond_Navoret_2015}
{\sc P.~Degond and L.~Navoret}, {\em A multi-layer model for self-propelled
  disks interacting through alignment and volume exclusion}, Math. Models
  Methods Appl. Sci., 25 (2015), pp.~2439--2475.

\bibitem{degond2010}
{\sc P.~Degond and T.~Yang}, {\em Diffusion in a continuum model of
  self-propelled particles with alignment interaction}, Math. Models Methods
  Appl. Sci., 20 (2010), pp.~1459--1490.

\bibitem{Dworkin1996}
{\sc M.~Dworkin}, {\em Recent advances in the social and developmental biology
  of the myxobacteria}, Microbiol. Rev., 60 (1996), pp.~70--102.

\bibitem{Eftimie2012}
{\sc R.~Eftimie}, {\em {Hyperbolic and kinetic models for self-organized
  biological aggregations and movement: A brief review}}, J. Math. Biol., 65
  (2012), pp.~35--75.

\bibitem{Frouvelle2012}
{\sc A.~Frouvelle}, {\em A continuum model for alignment of self-propelled
  particles with anisotropy and density-dependent parameters}, Math. Models
  Methods Appl. Sci., 22 (2012), p.~1250011.

\bibitem{Grossmann2016}
{\sc R.~Gro{\ss}mann, F.~Peruani, and M.~B{\"{a}}r}, {\em Mesoscale pattern
  formation of self-propelled rods with velocity reversal}, Phys. Rev. E, 94
  (2016), p.~050602(R).

\bibitem{gyllenberg1990nonlinear}
{\sc M.~Gyllenberg and G.~F. Webb}, {\em A nonlinear structured population
  model of tumor growth with quiescence}, J. Math. Biol., 28 (1990),
  pp.~671--694.

\bibitem{Igoshin2001}
{\sc O.~Igoshin, A.~Mogilner, R.~Welch, D.~Kaiser, and G.~Oster}, {\em Pattern
  formation and traveling waves in myxobacteria: theory and modeling}, Proc.
  Natl. Acad. Sci. USA, 98 (2001), pp.~14913--14918.

\bibitem{Igoshin2004b}
{\sc O.~Igoshin, J.~Neu, and G.~Oster}, {\em Developmental waves in
  myxobacteria: A distinctive pattern formation mechanism}, Phys. Rev. E, 70
  (2004), pp.~1--11.

\bibitem{Jelsbak2003}
{\sc L.~Jelsbak and L.~S{\o}gaard-Andersen}, {\em Cell behavior and cell-cell
  communication during fruiting body morphogenesis in {M}yxococcus xanthus}, J.
  Microbiol. Methods, 55 (2003), pp.~829--839.

\bibitem{Kim1990}
{\sc S.~Kim and D.~Kaiser}, {\em C-factor: A cell-cell signaling protein
  required for fruiting body morphogenesis of {M}. xanthus}, Cell, 61 (1990),
  pp.~19--26.

\bibitem{Lancon2002}
{\sc P.~Lancon, G.~Batrouni, L.~Lobry, and N.~Ostrowsky}, {\em {Brownian walker
  in a confined geometry leading to a space-dependent diffusion coefficient}},
  Phys. A, 304 (2002), p.~65.

\bibitem{Lutscher2002}
{\sc F.~Lutscher and A.~Stevens}, {\em {Emerging patterns in a hyperbolic model
  for locally interacting cell systems}}, J. Nonlinear Sci., 12 (2002),
  pp.~619--640.

\bibitem{Manhart2016}
{\sc A.~Manhart and A.~Mogilner}, {\em Agent-based modeling in cell biology},
  Mol. Biol. Cell., 27 (2016), pp.~3379--84.

\bibitem{Mogilner1999}
{\sc A.~Mogilner and L.~Edelstein-Keshet}, {\em A non-local model for a swarm},
  J. Math. Biol., 38 (1999), pp.~534--570.

\bibitem{Motsch_Navoret_MMS11}
{\sc S.~Motsch and L.~Navoret}, {\em Numerical simulations of a nonconvervative
  hyperbolic system with geometric constraints describing swarming behavior},
  Multiscale Model. Simul., 9 (2011), pp.~1253--1275.

\bibitem{Ngo2012}
{\sc S.~Ngo, F.~Ginelli, and H.~Chat{\'e}}, {\em Competing ferromagnetic and
  nematic alignment in self-propelled polar particles}, Phys. Rev. E, 86
  (2012), p.~050101.

\bibitem{pakdaman2009dynamics}
{\sc K.~Pakdaman, B.~Perthame, and D.~Salort}, {\em Dynamics of a structured
  neuron population}, Nonlinearity, 23 (2009), p.~55.

\bibitem{Peruani2011}
{\sc F.~Peruani, F.~Ginelli, M.~B{\"a}r, and H.~Chat{\'e}}, {\em Polar vs.
  apolar alignment in systems of polar self-propelled particles}, J. Phys.
  Conf. Ser., 297 (2011), p.~012014.

\bibitem{reichenbach2001}
{\sc H.~Reichenbach}, {\em Myxobacteria, producers of novel bioactive
  substances}, J. Ind. Microbiol. Biotechnol., 27 (2001), pp.~149--156.

\bibitem{Sager1994}
{\sc B.~Sager and D.~Kaiser}, {\em Intercellular {C}-signaling and the
  traveling waves of {M}yxococcus}, Genes Dev., 8 (1994), pp.~2793--2804.

\bibitem{scheel2017}
{\sc A.~Scheel and A.~Stevens}, {\em Wavenumber selection in coupled transport
  equations}, Journal of Mathematical Biology, 75 (2017), pp.~1047--1073.

\bibitem{shaffer1975}
{\sc B.~Shaffer}, {\em Secretion of cyclic {AMP} induced by cyclic {AMP} in the
  cellular slime mould dictyostelium discoideum}, Nature, 255 (1975),
  pp.~549--552.

\bibitem{Shimkets1990}
{\sc L.~J. Shimkets}, {\em Social and developmental biology of the
  myxobacteria.}, Microbiol. Rev., 54 (1990), pp.~473--501.

\bibitem{Sliusarenko2006}
{\sc O.~Sliusarenko, J.~Neu, D.~R. Zusman, and G.~Oster}, {\em Accordion waves
  in {M}yxococcus xanthus}, Proc. Natl. Acad. Sci. U S A, 103 (2006),
  pp.~1534--9.

\bibitem{Toner_Tu_1995}
{\sc J.~Toner and Y.~Tu}, {\em Long-range order in a two-dimensional dynamical
  {XY} model: how birds fly together}, Phys. Rev. Lett., 75 (1995),
  pp.~4326--4329.

\bibitem{Vicsek_Zafeiris_PhysRep12}
{\sc T.~Vicsek and A.~Zafeiris}, {\em Collective motion}, Phys. Rep., 517
  (2012), pp.~71--140.

\bibitem{Welch2001}
{\sc R.~Welch and D.~Kaiser}, {\em Cell behavior in traveling wave patterns of
  myxobacteria}, Proc. Natl. Acad. Sci. U S A, 98 (2001), pp.~14907--14912.

\end{thebibliography}

\end{document}